\documentclass{birkjour}
\usepackage{cite}
\usepackage{MnSymbol}
\usepackage{hyperref}
\newtheorem{thm}{Theorem}

\newtheorem{cor}{Corollary}

\newtheorem{lem}{Lemma}

\newtheorem{prop}{Proposition}

\theoremstyle{definition}
\newtheorem{defn}{Definition}

\theoremstyle{remark}
\newtheorem{rem}{Remark}

\begin{document}
\title{Quantum optimal transport with quantum channels}
\author{Giacomo De Palma}
\address{%
Massachusetts Institute of Technology\\
Research Laboratory of Electronics and Department of Mechanical Engineering\\
77 Massachusetts Avenue\\
02139 Cambridge MA\\
USA}

\email{gdepalma@mit.edu}

\author{Dario Trevisan}
\address{%
Universit\`a degli Studi di Pisa\\
Dipartimento di Matematica\\
56127 Pisa\\
Italy}

\email{dario.trevisan@unipi.it}

\begin{abstract}
We propose a new generalization to quantum states of the Wasserstein distance, which is a fundamental distance between probability distributions given by the minimization of a transport cost.
Our proposal is the first where the transport plans between quantum states are in natural correspondence with quantum channels, such that the transport can be interpreted as a physical operation on the system.
Our main result is the proof of a modified triangle inequality for our transport distance.
We also prove that the distance between a quantum state and itself is intimately connected with the Wigner-Yanase metric on the manifold of quantum states.
We then specialize to quantum Gaussian systems, which provide the mathematical model for the electromagnetic radiation in the quantum regime.
We prove that the noiseless quantum Gaussian attenuators and amplifiers are the optimal transport plans between thermal quantum Gaussian states, and that our distance recovers the classical Wasserstein distance in the semiclassical limit.
\end{abstract}

\maketitle

\tableofcontents

\section{Introduction}
Optimal transport theory \cite{villani2008optimal, ambrosio2008gradient, santambrogio2015book} is the study of the optimal transportation of resources and has now become a fundamental part of functional analysis with continuously growing applications. Indeed, optimal transport theory provides novel tools to tackle fundamental problems such as:
\begin{itemize}
\item the study partial differential equations, by interpreting many evolution equations as gradient flows with respect to transport-induced metrics \cite{otto2001porous};
\item geometric analysis,  with quantitative isoperimetric inequalities \cite{figalli2010isoperimetric} and synthetic notions of Ricci curvature bounds \cite{sturm2005ricci, lott2009ricci};
\item stochastic analysis in infinite dimensions \cite{ambrosio2009fokker};
\item random combinatorial optimization problems \cite{caracciolo2014scaling};
\item statistics and machine learning \cite{cuturi2019computational}.
\end{itemize}

In its original formulation \cite{monge}, the optimal transport problem looks for the cheapest way to transport a source mass distribution onto a target one. In mathematical terms, source and target are modelled via probability measures $\mu$, $\nu$ in $\mathbb{R}^n$, and the cost of transporting a unit of mass from a position $x$ to a position $y$ is a given function $c(y,x)$, whose most common choice is of the form $c(y,x) = |y-x|^p$ for some $p>0$. The assignment from the source $\mu$ to the target $\nu$ can be modelled as a ``transport map'' $f: \mathbb{R}^n \to \mathbb{R}^n$ such that for every open set $A \subseteq \mathbb{R}^n$
\begin{equation}\label{eq:deff}
\int_A \mathrm{d} \nu(x) =\int_{f^{-1}(A)} \mathrm{d} \mu(x)\,,
\end{equation}
where $f^{-1}(A)$ is the preimage of $A$. The overall transportation cost associated to $f$ is then
\begin{equation}
C(f) = \int_{\mathbb{R}^n} c(f(x), x)\,\mathrm{d} \mu(x)\,,
\end{equation}
and any minimizer $f^*$ of such a cost is called an optimal transport map.
The existence of a transport map $f$ satisfying \eqref{eq:deff} is in general not guaranteed.
For example, $f$ does not exist if $\mu$ and $\nu$ have support on finite sets with different number of points. This crucial issue was solved by relaxing the problem \cite{kantorovich1942transport}, introducing the so-called ``transport plans'' or couplings, \emph{i.e.}, probability measures $\pi$ on the product $\mathbb{R}^n\times\mathbb{R}^n$ such that their first and second marginal laws are respectively $\mu$ and $\nu$.
We denote with $\mathcal{C}(\mu,\nu)$ the set of such couplings.
Any transport map $f$ induces the coupling $\pi(y,x)= \delta_{y=f(x)}\,\mu(x)$,
but other coupling always exist,  \emph{e.g.}, the product measure $\pi=\nu\otimes \mu$.

From the disintegration theorem \cite[Volume II, Section 10.6]{bogachev2008measure}, we can associate to each coupling $\pi$ a stochastic map $\phi$ that assigns to each $x$ in the support of $\mu$ the probability measure $\phi(\cdot |x)$ on $\mathbb{R}^n$ such that for any measurable function $\psi:\mathbb{R}^n\times\mathbb{R}^n\to\mathbb{R}$
\begin{equation}\label{eq:defpi|}
\int_{\mathbb{R}^n\times\mathbb{R}^n} \psi(y,x)\,\mathrm{d}\phi(y|x)\,\mathrm{d}\mu(x) = \int_{\mathbb{R}^n\times\mathbb{R}^n} \psi(y,x)\,\mathrm{d}\pi(y,x)\,,
\end{equation}
\emph{i.e.}, $\mathrm{d}\phi(y|x)$ is the conditional probability distribution of $y$ given $x$ induced by $\pi$.
The stochastic map $\phi$ sends a probability measure $\rho$ defined on the support of $\mu$ to the probability measure on $\mathbb{R}^n$
\begin{equation}\label{eq:actphi}
\phi(\rho) = \int_{\mathbb{R}^n}\phi(\cdot|x)\,\mathrm{d}\rho(x)\,.
\end{equation}
We notice that from \eqref{eq:defpi|}, $\phi(\mu) = \nu$.

The cost associated to the coupling $\pi$ is
\begin{equation}\label{eq:defCpi}
C(\pi) = \int_{\mathbb{R}^n \times \mathbb{R}^n} c(y,x)\, \mathrm{d}\pi(y,x)\,,
\end{equation}
and the optimal transport cost is given by
\begin{equation}
{W(\mu,\nu)}^2 = \inf_{\pi\in\mathcal{C}(\mu,\nu)}C(\pi)\,.
\end{equation}
The optimal transport problem is now relaxed to a linear optimization problem that under mild regularity assumptions always admits a solution. In special cases, one can a posteriori prove that an optimal coupling is in fact a transport map, the most notable case being that of $c(y,x) = |y-x|^2$, when $\mu$ is absolutely continuous \cite{brenier1987polar}. In fact, with the same cost, the square root  of the associated optimal transport cost provides a distance on the space of probability measures, commonly denoted $W(\mu, \nu)$ and called Wasserstein distance, which induces a Riemannian metric on the manifold of probability measures on $\mathbb{R}^n$ and whose geometric properties play an essential role in many of the applications mentioned above \cite{otto2001porous,figalli2010isoperimetric, sturm2005ricci, lott2009ricci, ambrosio2009fokker, caracciolo2014scaling}.

\subsection{Our contribution}
There have been two recent proposals to generalize the Wasserstein distance to the quantum setting.
The first proposal by Carlen, Maas, Datta and Rouz\'e \cite{carlen_analog_2014,carlen_gradient_2017,carlen2020non,rouze2019concentration,datta2020relating,chen2018matrix,chen2018wasserstein} is built on the equivalent dynamical definition of the Wasserstein distance provided by Benamou and Brenier \cite{benamou2000computational}, which assigns a length to each path of probability measures that connects the source with the target.
The key property of this proposal is that the resulting quantum distance is induced by a Riemannian metric on the manifold of quantum states, and the quantum generalization of the heat semigroup is the gradient flow of the von Neumann entropy with respect to this metric.
This quantum generalization of the Wasserstein distance has been shown to be intimately linked to both entropy and Fisher information \cite{datta2020relating}, and has led to determine the rate of convergence of the quantum Ornstein-Uhlenbeck semigroup \cite{carlen_gradient_2017,huber2018conditional}.
The second proposal by Golse, Mouhot, Paul and Caglioti \cite{golse2016mean,caglioti2018towards,golse2018quantum,golse2017schrodinger,golse2018wave,caglioti2019quantum} arose in the context of the study of the semiclassical limit of quantum mechanics and is built on the definition of the Wasserstein distance through couplings \eqref{eq:defCpi}.
This distance was the key element to prove that the mean-field limit of quantum mechanics is uniform in the semiclassical limit \cite{golse2016mean}, and has been employed as a cost function to train the quantum counterpart of deep generative adversarial networks \cite{lloyd2018quantum,chakrabarti2019quantum}.
In the following, we will refer to this distance as the GMPC distance.

A fundamental property of the classical transport plans is that they are in one-to-one correspondence with stochastic maps, and this correspondence provides the operational interpretation of the Wasserstein distance as the distance associated to a physical operation that is performed on the system of interest.
The mathematical model for the physical operations that can be performed on a quantum system are the quantum channels, which are the completely-positive and trace-preserving linear maps on the set of trace-class operators on the Hilbert space of the quantum system, and are the quantum counterpart of the stochastic maps \cite{wilde2017quantum,breuer2007theory,nielsen2010quantum,holevo2019quantum}.
We propose a new quantum generalization of the Wasserstein distance that builds on the GMPC distance and has the key property that the associated set of quantum transport plans is in a one-to-one correspondence with the set of quantum channels.
Our proposal is the first that has this property, which allows for the operational interpretation of quantum transport plans as physical operations performed on the quantum system.

Our main result is that our quantum generalization of the Wasserstein distance satisfies a modified triangle inequality (\autoref{thm:triangle}), whose validity for the GMPC distance is not known.
Our distance shares with the GMPC distance the peculiar property of being nonzero even for coinciding quantum states, and we prove that our distance between a quantum state and itself is intimately connected to the Wigner-Yanase metric on the set of quantum states \cite{gibilisco2003wigner,bengtsson2017geometry}.
We then focus on quantum Gaussian systems, which provide the mathematical model for the electromagnetic radiation in the quantum regime.
Quantum Gaussian systems play a central role in quantum information, since photons traveling through optical fibers provide the main platform for quantum key distribution and one of the most promising platforms for quantum computation \cite{braunstein2005quantum,weedbrook2012gaussian,schuld2018supervised}.
We prove that the optimal transport plans between thermal quantum Gaussian states are noiseless quantum Gaussian attenuators or amplifiers, which model the attenuation of electromagnetic signals traveling through optical fibers and their optimal amplification, respectively (\autoref{prop:thermal}).
We also show that the distance between generic states is convex with respect to mixing with a beamsplitter (\autoref{thm:ou}) and subadditive with respect to the addition of classical noise (\autoref{thm:heat}).
Moreover, we prove that our distance between a state of a quantum Gaussian system and itself satisfies a Stam inequality (\autoref{thm:Stam}).
Finally, we prove that our distance recovers the classical Wasserstein distance in the semiclassical limit.
Specifically, our distance is lower bounded by the Wasserstein distance between the Husimi Q representations of the quantum states (\autoref{thm:classical2}), and if the quantum states are semiclassical, it is also upper bounded by the Wasserstein distance between their Glauber-Sudarshan P representations (\autoref{thm:classical1}).

The paper is structured as follows.
In \autoref{sec:plans}, we present our definition of quantum transport plans and show that they are in a one-to-one correspondence with quantum channels.
In \autoref{sec:W}, we define our quantum Wasserstein distance and prove that it satisfies a modified triangle inequality, and in \autoref{sec:WY} we show the connection with the Wigner-Yanase metric.
In \autoref{sec:intro} we introduce quantum Gaussian systems, in \autoref{sec:WG} we determine the optimal transport plans between thermal quantum Gaussian states, and in \autoref{sec:classical} we study the semiclassical limit of the quantum Wasserstein distance.
We conclude in \autoref{sec:concl}.
We prove in \autoref{sec:continuity} some results that are needed to deal with Hilbert spaces with infinite dimension and unbounded operators, and \autoref{sec:app} contains the proof of an auxilary lemma.

\section{Quantum transport plans and quantum couplings}\label{sec:plans}
\subsection{Preliminaries}
Let $\mathcal{H}$ be a separable Hilbert space and let $\mathcal{L}(\mathcal{H})$ be the set of linear operators on $\mathcal{H}$.
Let $\mathcal{B}(\mathcal{H})$ be the set of bounded linear operators on $\mathcal{H}$, let
\begin{equation}
\mathcal{T}_1(\mathcal{H}) = \left\{X\in\mathcal{B}(\mathcal{H}):\mathrm{Tr}_{\mathcal{H}}\sqrt{X^\dag X}<\infty\right\}
\end{equation}
be the set of trace-class operators on $\mathcal{H}$, and let
\begin{equation}
\mathcal{S}(\mathcal{H}) = \left\{\rho\in\mathcal{T}_1(\mathcal{H}):\rho\ge0,\;\mathrm{Tr}_{\mathcal{H}}\rho=1\right\}
\end{equation}
be the set of the quantum states of $\mathcal{H}$, which are the quantum counterpart of the classical probability measures.
The quantum counterparts of the stochastic maps are the quantum channels, which are the completely-positive and trace-preserving linear maps on $\mathcal{T}_1(\mathcal{H})$.
We recall that a linear map $\Phi$ is positive if it preserves the set of nonnegative operators in $\mathcal{T}_1(\mathcal{H})$ and completely positive if the linear map $\Phi\otimes\mathbb{I}_n$ acting on $\mathcal{T}_1\left(\mathcal{H}\otimes\mathbb{C}^n\right)$ is positive for any $n\in\mathbb{N}$ \cite{holevo2019quantum}.
Quantum channels preserve the set of quantum states even when they are applied only to a subsystem, and all the linear maps with this property are quantum channels.

Let
\begin{equation}
\mathcal{T}_2(\mathcal{H}) = \left\{X\in\mathcal{B}(\mathcal{H}):\left\|X\right\|_2^2 = \mathrm{Tr}_{\mathcal{H}}\left[X^\dag X\right]<\infty\right\}
\end{equation}
be the set of the Hilbert-Schmidt operators on $\mathcal{H}$.
$\mathcal{T}_2(\mathcal{H})$ is a Hilbert space if equipped with the hermitian product
\begin{equation}
\langle\langle X||Y\rangle\rangle = \mathrm{Tr}_{\mathcal{H}}\left[X^\dag\,Y\right]\,,\qquad X,\,Y\in\mathcal{T}_2(\mathcal{H})\,.
\end{equation}
Let $\mathcal{H}^*$ be the Hilbert space of continuous linear functionals on $\mathcal{H}$.
\begin{defn}[transpose]
For any $X\in\mathcal{L}(\mathcal{H})$, let $X^T$ be the linear operator on $\mathcal{H}^*$ given by
\begin{equation}
X^T\langle\phi| = \langle\phi|X\,,\qquad\langle\phi|\in\mathcal{H}^*\,.
\end{equation}
\end{defn}
\begin{prop}[\cite{royer1991wigner,d2000bell}]
$\mathcal{H}\otimes\mathcal{H}^*$ is canonically isometric to $\mathcal{T}_2(\mathcal{H})$ through the linear extension of the map
\begin{equation}
|\psi\rangle\otimes\langle\phi|\mapsto |\psi\rangle\langle\phi|\,,\qquad |\psi\rangle\in\mathcal{H},\;\langle\phi|\in\mathcal{H}^*\,.
\end{equation}
\end{prop}
We will make an extensive use of the following properties:
\begin{lem}[\cite{royer1991wigner,d2000bell}]\label{lem:AXB}
For any $X\in\mathcal{T}_2(\mathcal{H})$ and any $A,\,B\in\mathcal{L}(\mathcal{H})$ such that $A\,X\,B\in\mathcal{T}_2(\mathcal{H})$ we have
\begin{equation}
\left(A\otimes B^T\right)\|X\rangle\rangle = \|A\,X\,B\rangle\rangle\,.
\end{equation}
\end{lem}
\begin{lem}\label{lem:Tr}
For any $X\in\mathcal{T}_2(\mathcal{H})$ we have
\begin{equation}
\mathrm{Tr}_{\mathcal{H}^*}||X\rangle\rangle\langle\langle X|| = X\,X^\dag\,,\qquad \mathrm{Tr}_{\mathcal{H}}||X\rangle\rangle\langle\langle X|| = \left(X^\dag X\right)^T\,.
\end{equation}
\end{lem}
A purification of a quantum state $\rho\in\mathcal{S}(\mathcal{H})$ is a pure quantum state $\gamma\in\mathcal{S}(\mathcal{H}\otimes\mathcal{K})$, where $\mathcal{K}$ is a separable Hilbert space, such that
\begin{equation}
\mathrm{Tr}_\mathcal{K}\,\gamma = \rho\,.
\end{equation}
The Hilbert space $\mathcal{K}$ can always be chosen to be $\mathcal{H}^*$:
\begin{defn}[canonical purification \cite{holevo2019quantum}]
The canonical purification of a quantum state $\rho\in\mathcal{S}(\mathcal{H})$ is $\left\|\sqrt{\rho}\right\rrangle\left\llangle\sqrt{\rho}\right\|\in\mathcal{S}\left(\mathcal{H}\otimes\mathcal{H}^*\right)$.
\end{defn}

\subsection{Quantum transport}
We start defining our notion of quantum transport plan as the quantum counterpart of the classical stochastic map $\phi$ defined in \eqref{eq:defpi|}.
\begin{defn}[quantum transport plan]\label{def:QT}
For any $\rho,\,\sigma\in\mathcal{S}(\mathcal{H})$, the set $\mathcal{M}(\rho,\sigma)$ of quantum transport plans from $\rho$ to $\sigma$ is the set of the quantum channels $\Phi:\mathcal{T}_1(\mathrm{supp}\,\rho)\to\mathcal{T}_1(\mathcal{H})$ such that $\Phi(\rho) = \sigma$.
\end{defn}
GMPC associate to any $\rho,\,\sigma\in\mathcal{S}(\mathcal{H})$ the set of quantum couplings
\begin{equation}
\mathcal{C}_{\textnormal{GMPC}}(\rho,\sigma) = \left\{\Pi\in\mathcal{S}(\mathcal{H}_2\otimes\mathcal{H}_1):\mathrm{Tr}_{\mathcal{H}_2}\Pi = \rho\,,\,\mathrm{Tr}_{\mathcal{H}_1}\Pi = \sigma\right\}\,,
\end{equation}
where $\mathcal{H}_{1,2}$ are two copies of $\mathcal{H}$ \cite{golse2016mean}.
With this definition, there is no straightforward way to associate a quantum transport plan to a quantum coupling.

We propose a new definition of quantum coupling that admits a one-to-one correspondence with quantum transport plans.
First, we associate to any quantum transport plan $\Phi\in\mathcal{M}(\rho,\sigma)$ the quantum state of $\mathcal{H}\otimes\mathcal{H}^*$
\begin{equation}\label{eq:defPi}
\Pi_\Phi = \left(\Phi\otimes\mathbb{I}_{\mathcal{T}_1(\mathcal{H}^*)}\right) \left(\left\|\sqrt{\rho}\right\rrangle\left\llangle\sqrt{\rho}\right\|\right)\,.
\end{equation}
We get from \autoref{lem:Tr}
\begin{equation}
\mathrm{Tr}_{\mathcal{H}^*}\left[\left\|\sqrt{\rho}\right\rrangle\left\llangle\sqrt{\rho}\right\|\right] = \rho\,,\qquad \mathrm{Tr}_{\mathcal{H}}\left[\left\|\sqrt{\rho}\right\rrangle\left\llangle\sqrt{\rho}\right\|\right] = \rho^T\,,
\end{equation}
therefore
\begin{equation}\label{eq:marginals}
\mathrm{Tr}_{\mathcal{H}}\Pi_\Phi = \rho^T\,,\qquad \mathrm{Tr}_{\mathcal{H}^*}\Pi_\Phi = \sigma\,.
\end{equation}
\begin{defn}[quantum coupling]\label{defn:coupling}
In view of \eqref{eq:marginals}, we associate to any $\rho,\,\sigma\in\mathcal{S}(\mathcal{H})$ the set of quantum couplings
\begin{equation}
\mathcal{C}(\rho,\sigma) = \left\{\Pi\in\mathcal{S}(\mathcal{H}\otimes\mathcal{H}^*):\mathrm{Tr}_{\mathcal{H}}\Pi = \rho^T\,,\,\mathrm{Tr}_{\mathcal{H}^*}\Pi = \sigma\right\}\,.
\end{equation}
\end{defn}
\begin{rem}
If the partial transpose on $\mathcal{H}^*$ of $\Pi\in\mathcal{C}(\rho,\sigma)$ is positive semi-definite, it belongs to $\mathcal{C}_{\textnormal{GMPC}}(\rho,\sigma)$.
Moreover, if the partial transpose on $\mathcal{H}_1$ of $\Pi'\in\mathcal{C}_{\textnormal{GMPC}}(\rho,\sigma)$ is positive semi-definite, it belongs to $\mathcal{C}(\rho,\sigma)$.
Therefore, the quantum couplings of $\mathcal{C}(\rho,\sigma)$ with positive partial transpose are in a one-to-one correspondence with the quantum couplings of $\mathcal{C}_{\textnormal{GMPC}}(\rho,\sigma)$ with positive partial transpose.
On the contrary, the couplings of $\mathcal{C}(\rho,\sigma)$ with non-positive partial transpose do not have any counterpart in $\mathcal{C}_{\textnormal{GMPC}}(\rho,\sigma)$, and vice-versa.
\end{rem}
\begin{rem}
A definition of quantum coupling very similar to \autoref{defn:coupling} has been proposed in \cite{winter2016tight}, where a coupling between $\rho$ and $\sigma$ is defined as a quantum state of $\mathcal{H}^{\otimes2}$ with marginals $\rho$ and $\sigma^T$, where $\sigma^T$ is the quantum state of $\mathcal{H}$ whose density matrix is the transpose of the density matrix of $\sigma$ in a given basis.
\end{rem}
\begin{rem}\label{rem:I}
The quantum coupling associated to the trivial transport plan of $\rho$ onto itself is the canonical purification of $\rho$:
\begin{equation}
\Pi_{\mathbb{I}} = \left\|\sqrt{\rho}\right\rrangle\left\llangle\sqrt{\rho}\right\|\,.
\end{equation}
Unless $\rho$ is a pure state, the partial transpose of this coupling is not positive.
Therefore, this coupling does not have any counterpart in $\mathcal{C}_{\textnormal{GMPC}}(\rho,\rho)$.
\end{rem}

The following \autoref{prop:CM} proves that our quantum couplings are in a one-to-one correspondence with the quantum transport plans.
\begin{prop}\label{prop:CM}
For any $\rho,\,\sigma\in\mathcal{S}(\mathcal{H})$, the map $\Phi\mapsto\Pi_\Phi$ defined by \eqref{eq:defPi} is a bijection between $\mathcal{M}(\rho,\sigma)$ and $\mathcal{C}(\rho,\sigma)$.
\begin{proof}
Let us prove that the map is surjective.
Let $\Pi\in\mathcal{C}(\rho,\sigma)$, and let
\begin{equation}\label{eq:condA}
\Pi = \sum_{n=0}^\infty p_n\,||A_n\rangle\rangle\langle\langle A_n||
\end{equation}
be an orthonormal eigendecomposition of $\Pi$ as a linear operator on the Hilbert space $\mathcal{T}_2(\mathcal{H})$, where the series converges in the trace norm.
Since $\Pi\in\mathcal{C}(\rho,\sigma)$, we get from \autoref{lem:Tr}
\begin{equation}\label{eq:rhosum}
\sum_{n=0}^\infty p_n\, A_n^\dag\,A_n = \rho\,,\qquad\sum_{n=0}^\infty p_n\,A_n\,A_n^\dag = \sigma\,,
\end{equation}
where both series converge in the trace norm.
We define for any $X \in \mathcal{T}_1(\mathrm{supp}\,\rho)$
\begin{equation}\label{eq:defPhiPi}
\Phi_\Pi(X) = \sum_{n=0}^\infty p_n\,A_n\,\rho^{-\frac{1}{2}}\,X\,\rho^{-\frac{1}{2}}\,A_n^\dag\in\mathcal{T}_1(\mathcal{H})\,.
\end{equation}
The series in \eqref{eq:defPhiPi} converges in the trace norm since for any $N\in\mathbb{N}$
\begin{align}
&\sum_{n=0}^N p_n\left\|A_n\,\rho^{-\frac{1}{2}}\,X\,\rho^{-\frac{1}{2}}\,A_n^\dag\right\|_1 \le \sum_{n=0}^N p_n\,\mathrm{Tr}\left[A_n\,\rho^{-\frac{1}{2}}\left|X\right|\rho^{-\frac{1}{2}}\,A_n^\dag\right]\nonumber\\
&= \mathrm{Tr}\left[\left|X\right|\sum_{n=0}^N p_n\,\rho^{-\frac{1}{2}}\,A_n^\dag A_n\,\rho^{-\frac{1}{2}}\right] \le \left\|X\right\|_1\,,
\end{align}
where the last inequality follows from \eqref{eq:rhosum}.
The map $\Phi_\Pi$ defined in \eqref{eq:defPhiPi} is linear and completely positive, and from \eqref{eq:condA} it is trace preserving and satisfies $\Phi_\Pi(\rho)=\sigma$, hence $\Phi_\Pi\in\mathcal{M}(\rho,\sigma)$.
Moreover, if $\Pi$ has degenerate spectrum, $\Phi_\Pi$ does not depend on the choice of the eigenvectors of $\Pi$ in \eqref{eq:condA}.
Indeed, for any $X\in\mathcal{T}_1(\mathrm{supp}\,\rho)$ such that $\left\|\rho^{-\frac{1}{2}}\,X\,\rho^{-\frac{1}{2}}\right\|_\infty<\infty$ we have
\begin{equation}\label{eq:Phitr}
\Phi_\Pi(X)= \mathrm{Tr}_{\mathcal{H}^*}\left[\left(\mathbb{I}_{\mathcal{H}}\otimes\left(\rho^{-\frac{1}{2}}\,X\,\rho^{-\frac{1}{2}}\right)^T\right)\Pi\right]\,.
\end{equation}
Since $\Phi_\Pi$ is completely positive and trace preserving, it is continuous with respect to the trace norm, hence $\Phi_\Pi(X)$ does not depend on the choice of the eigenvectors for any $X\in\mathcal{T}_1(\mathrm{supp}\,\rho)$.
With the help of \autoref{lem:AXB} we have
\begin{align}
\Pi_{\Phi_\Pi} &= \sum_{n=0}^\infty p_n\left(A_n\,\rho^{-\frac{1}{2}}\otimes\mathbb{I}_{\mathcal{H^*}}\right) \left\|\sqrt{\rho}\right\rrangle\left\llangle\sqrt{\rho}\right\| \left(\rho^{-\frac{1}{2}}\,A_n^\dag\otimes\mathbb{I}_{\mathcal{H^*}}\right)\nonumber\\
& = \sum_{n=0}^\infty p_n\,\left\|A_n\right\rrangle\left\llangle A_n\right\| = \Pi\,,
\end{align}
hence the map $\Phi\mapsto\Pi_\Phi$ is surjective.

Let us show that it is also injective.
Let $\Phi,\,\Phi'\in\mathcal{M}(\rho,\sigma)$ such that $\Pi_\Phi = \Pi_{\Phi'} = \Pi$.
Recalling the definition \eqref{eq:defPi} of $\Pi_\Phi$, for any $X\in\mathcal{T}_1(\mathrm{supp}\,\rho)$ such that $\left\|\rho^{-\frac{1}{2}}\,X\,\rho^{-\frac{1}{2}}\right\|_\infty<\infty$ we have
\begin{align}
\Phi_\Pi(X) &= \Phi \left(\mathrm{Tr}_{\mathcal{H}^*}\left[\left(\mathbb{I}_{\mathcal{H}}\otimes\left(\rho^{-\frac{1}{2}}\,X\,\rho^{-\frac{1}{2}}\right)^T\right)
\left\|\sqrt{\rho}\right\rrangle\left\llangle\sqrt{\rho}\right\|\right]\right)\nonumber\\
&= \Phi \left(\mathrm{Tr}_{\mathcal{H}^*}\left[\left\|X\,\rho^{-\frac{1}{2}}\right\rrangle\left\llangle\rho^\frac{1}{2}\right\|\right]\right) = \Phi(X)\,,
\end{align}
where the second equality follows from \autoref{lem:AXB}.
Since $\Phi_\Pi$ is continuous with respect to the trace norm, we get $\Phi_\Pi(X)=\Phi(X)$ for any $X\in\mathcal{T}_1(\mathrm{supp}\,\rho)$, hence $\Phi_\Pi = \Phi$.
Analogously we get $\Phi_\Pi = \Phi'$, hence $\Phi=\Phi'$.
\end{proof}
\end{prop}

\section{Quantum transport cost and quantum Wasserstein distance}\label{sec:W}
Given a set $\{R_1,\,\ldots,\,R_N\}$ of self-adjoint operators on $\mathcal{H}$, which we call \emph{quadratures}, we propose the following operational definition of the transport cost associated to the transport plan $\Phi$ applied to the quantum state $\rho$.
We build $N$ copies of the quantum state $\Pi_\Phi$ defined in \eqref{eq:defPi}, and for each $i=1,\,\ldots,\,N$ we measure $R_i$ on the $\mathcal{H}$ subsystem and $R_i^T$ on the $\mathcal{H}^*$ subsystem of the $i$-th copy, getting the outcomes $r_i$ and $r_i'$, respectively.
We define the transport cost as the expectation value of $\sum_{i=1}^N\left(r_i-r_i'\right)^2$ over the above protocol.
This cost has a simple expression in terms of the quantum transport plan:
\begin{defn}[quantum transport cost]\label{def:cost}
Let $\Pi$ be a quantum state of $\mathcal{H}\otimes\mathcal{H}^*$ whose eigenvectors all belong to the domain of each $R_i\otimes\mathbb{I}_{\mathcal{H}^*} - \mathbb{I}_{\mathcal{H}}\otimes R_i^T$.
We define the cost associated to $\Pi$ as
\begin{equation}\label{eq:C}
C(\Pi) = \sum_{i=1}^N\mathrm{Tr}_{{\mathcal{H}\otimes\mathcal{H}^*}}\left[\left(R_i\otimes\mathbb{I}_{\mathcal{H}^*} - \mathbb{I}_{\mathcal{H}}\otimes R_i^T\right)\Pi\left(R_i\otimes\mathbb{I}_{\mathcal{H}^*} - \mathbb{I}_{\mathcal{H}}\otimes R_i^T\right)\right]\,.
\end{equation}
\end{defn}

\begin{rem}
Let $\Pi'$ be a quantum state of $\mathcal{H}^{\otimes 2}$ whose eigenvectors all belong to the domain of each $R_i\otimes\mathbb{I} - \mathbb{I}\otimes R_i$.
GMPC define the cost associated to $\Pi'$ as
\begin{equation}
C_{\textnormal{GMPC}}(\Pi') = \sum_{i=1}^N\mathrm{Tr}_{\mathcal{H}^{\otimes2}}\left[\left(R_i\otimes\mathbb{I} - \mathbb{I}\otimes R_i\right)\Pi'\left(R_i\otimes\mathbb{I} - \mathbb{I}\otimes R_i\right)\right]\,.
\end{equation}
Let $T_{\mathcal{H}^*}$ denote the partial transposition on $\mathcal{H}^*$.
Let $\Pi\in\mathcal{C}(\rho,\sigma)$ be such that $\Pi^{T_{\mathcal{H}^*}}\ge0$.
Then, $\Pi^{T_{\mathcal{H}^*}}\in\mathcal{C}_{\textnormal{GMPC}}(\rho,\sigma)$, and $C(\Pi) = C_{\textnormal{GMPC}}\left(\Pi^{T_{\mathcal{H}^*}}\right)$.
\end{rem}

\begin{rem}
We define the cost operator
\begin{equation}\label{eq:defC}
C = \sum_{i=1}^N\left(R_i\otimes\mathbb{I}_{\mathcal{H}^*} - \mathbb{I}_{\mathcal{H}}\otimes R_i^T\right)^2\,.
\end{equation}
Whenever $C\,\Pi\in\mathcal{T}_1(\mathcal{H}\otimes\mathcal{H}^*)$, the cost of $\Pi$ can also be expressed as
\begin{equation}
C(\Pi) = \mathrm{Tr}_{\mathcal{H}\otimes\mathcal{H}^*}\left[C\,\Pi\right]\,.
\end{equation}
\end{rem}

\begin{defn}[energy]\label{def:energy-main}
Let $A$ be a self-adjoint operator on $\mathcal{H}$ and let $\rho\in\mathcal{S}(\mathcal{H})$ be a quantum state with eigendecomposition $\rho = \sum_{i=0}^\infty  p_i\,|\psi_i\rangle\langle\psi_i|$.
We define the energy of $\rho$ with respect to $A$ as
\begin{equation}
E_A(\rho) = \sum_{i=0}^\infty p_i \left\|A|\psi_i\rangle\right\|^2 \in [0, \infty]
\end{equation}
if every $|\psi_i\rangle$ belongs to the domain of $A$, and $E_A(\rho) = \infty$ otherwise.

We define the energy of $\rho$ with respect to the quadratures $R_1,\,\ldots,\,R_N$ as
\begin{equation}\label{eq:defEn}
E(\rho) = E_{R_1}(\rho) + \ldots + E_{R_N}(\rho)\,.
\end{equation}
For the sake of a simpler notation, we will refer to \eqref{eq:defEn} simply as the energy of $\rho$.
\end{defn}

\begin{rem}\label{rem:en-pure}
If $\rho = |\psi \rangle \langle \psi |$ is pure, then $E_A(\rho) = \left\|A|\psi \rangle\right\|^2$ if $|\psi\rangle$ belongs to the domain of $A$, $E_A(\rho) = \infty$ otherwise.
\end{rem}

\begin{rem}\label{rem:en}
If $A \in \mathcal{B}(\mathcal{H})$, then $A\,\rho\,A\in\mathcal{T}_1(\mathcal{H})$ and
we have
\begin{equation}
\mathrm{Tr}_{\mathcal{H}}\left[A\,\rho\,A\right] = \sum_{i=0}^\infty p_i\left\|A|\psi_i\rangle\right\|^2  = E_A(\rho) <\infty\,.
\end{equation}
\end{rem}

\begin{lem}\label{lem:en}
 Let $\rho$ be a quantum state with finite energy.
Then, $R_i\,\rho\,R_i\in\mathcal{T}_1(\mathcal{H})$ for any $i=1,\,\ldots,\,N$.
\begin{proof}
Let
\begin{equation}
\rho = \sum_{n=0}^\infty p_n\,|\psi_n\rangle\langle\psi_n|
\end{equation}
be an eigendecomposition of $\rho$.
We have
\begin{equation}
\mathrm{Tr}_{\mathcal{H}}\left[R_i\,\rho\,R_i\right] = \sum_{n=0}^\infty p_n\left\|R_i|\psi_n\rangle\right\|^2 <\infty\,,
\end{equation}
and the claim follows.
\end{proof}
\end{lem}

\begin{prop}\label{prop:CE}
Let $\rho,\,\sigma\in\mathcal{S}(\mathcal{H})$ have finite energy.
Then, any plan $\Pi\in\mathcal{C}(\rho,\sigma)$ has finite cost.
\begin{proof}
See \autoref{sec:CEproof}.
\end{proof}
\end{prop}
\begin{defn}[swap transposition]
Let $\Gamma$ be an operator on $\mathcal{H}\otimes\mathcal{H}^*$.
$\Gamma^T$ is an operator on $\mathcal{H}^*\otimes\mathcal{H}$.
We define $\Gamma^{ST}$ to be the operator on $\mathcal{H}\otimes\mathcal{H}^*$ associated to $\Gamma^T$ through the canonical identification between $\mathcal{H}^*\otimes\mathcal{H}$ and $\mathcal{H}\otimes \mathcal{H}^*$.
\end{defn}
The swap transposition provides a canonical identification between $\mathcal{C}(\rho,\sigma)$ and $\mathcal{C}(\sigma,\rho)$:
\begin{prop}\label{prop:symmetry}
For any $\Pi\in\mathcal{C}(\rho,\sigma)$, we have $\Pi^{ST}\in\mathcal{C}(\sigma,\rho)$, and the two couplings have the same cost.
\end{prop}

As in the classical case, we define the square Wasserstein distance as the minimum transport cost:
\begin{defn}[quantum Wasserstein distance]\label{defn:D}
We define for any $\rho,\,\sigma\in\mathcal{S}(\mathcal{H})$
\begin{equation}\label{eq:defDd}
{D(\rho,\sigma)}^2 = \inf_{\Pi\in\mathcal{C}(\rho,\sigma)}C(\Pi)\,.
\end{equation}
\end{defn}
\begin{rem}
From \autoref{prop:symmetry}, $D(\rho,\sigma)=D(\sigma,\rho)$.
\end{rem}
\begin{rem}
\autoref{defn:D} is completely analogous to the GMPC definition of distance
\begin{equation}
{D_{\textnormal{GMPC}}(\rho,\sigma)}^2 = \inf_{\Pi'\in\mathcal{C}_{\textnormal{GMPC}}(\rho,\sigma)}C_{\textnormal{GMPC}}(\Pi')\,.
\end{equation}
\end{rem}

As in the classical case, the quantum Wasserstein distance is additive with respect to the tensor product:
\begin{prop}[additivity with respect to tensor product]\label{prop:additivity}
Let $\mathcal{H}_1$ and $\mathcal{H}_2$ be Hilbert spaces with quadratures $R_1^1,\,\ldots,\,R_{N_1}^1$ and $R_1^2,\,\ldots,\,R_{N_2}^2$, respectively, and let $R_1^1\otimes\mathbb{I}_{\mathcal{H}_2},\,\ldots,\,R_{N_1}^1\otimes\mathbb{I}_{\mathcal{H}_2},\,\mathbb{I}_{\mathcal{H}_1}\otimes R_1^2,\,\ldots,\,\mathbb{I}_{\mathcal{H}_1}\otimes R_{N_2}^2$ be the quadrature operators of $\mathcal{H} = \mathcal{H}_1\otimes\mathcal{H}_2$.
Then, for any $\rho_1,\,\sigma_1\in\mathcal{S}(\mathcal{H}_1)$ and any $\rho_2,\,\sigma_2\in\mathcal{S}(\mathcal{H}_2)$ with finite energy,
\begin{equation}
{D(\rho_1\otimes\rho_2,\,\sigma_1\otimes\sigma_2)}^2 = {D(\rho_1,\sigma_1)}^2 + {D(\rho_2,\sigma_2)}^2\,.
\end{equation}
\begin{proof}
We have for any $\Pi\in\mathcal{C}(\rho_1\otimes\rho_2,\,\sigma_1\otimes\sigma_2)$ with finite cost
\begin{align}
C(\Pi) &= \sum_{i=1}^{N_1} E_{R_i^1\otimes\mathbb{I}_{\mathcal{H}_2}\otimes\mathbb{I}_{\mathcal{H}^*} - \mathbb{I}_{\mathcal{H}}\otimes {R_i^1}^T\otimes\mathbb{I}_{\mathcal{H}_2^*}}(\Pi)\nonumber\\
&\phantom{=} + \sum_{i=1}^{N_2} E_{\mathbb{I}_{\mathcal{H}_1}\otimes R_i^2\otimes\mathbb{I}_{\mathcal{H}^*} - \mathbb{I}_{\mathcal{H}}\otimes \mathbb{I}_{\mathcal{H}_1^*}\otimes {R_i^2}^T}(\Pi)\nonumber\\
&= C_1\left(\mathrm{Tr}_{\mathcal{H}_2\otimes\mathcal{H}_2^*}\Pi\right) + C_2\left(\mathrm{Tr}_{\mathcal{H}_1\otimes\mathcal{H}_1^*}\Pi\right) \ge {D(\rho_1,\sigma_1)}^2 + {D(\rho_2,\sigma_2)}^2\,,
\end{align}
where we have used that $\mathrm{Tr}_{\mathcal{H}_2\otimes\mathcal{H}_2^*}\Pi\in\mathcal{C}(\rho_1,\sigma_1)$ and $\mathrm{Tr}_{\mathcal{H}_1\otimes\mathcal{H}_1^*}\Pi\in\mathcal{C}(\rho_2,\sigma_2)$.

Conversely, for any $\Pi_1\in\mathcal{C}(\rho_1,\sigma_1)$ and $\Pi_2\in\mathcal{C}(\rho_2,\sigma_2)$, we have that $\Pi=\Pi_1\otimes\Pi_2\in\mathcal{C}(\rho_1\otimes\rho_2,\,\sigma_1\otimes\sigma_2)$, hence
\begin{equation}\label{eq:Pi12}
{D(\rho_1\otimes\rho_2,\,\sigma_1\otimes\sigma_2)}^2 \le C(\Pi) = C_1(\Pi_1) + C_2(\Pi_2)\,.
\end{equation}
Taking the infimum of the right-hand side of \eqref{eq:Pi12} over $\Pi_1\in\mathcal{C}(\rho_1,\sigma_1)$ and $\Pi_2\in\mathcal{C}(\rho_2,\sigma_2)$ we get
\begin{equation}
{D(\rho_1\otimes\rho_2,\,\sigma_1\otimes\sigma_2)}^2 \le {D(\rho_1,\sigma_1)}^2 + {D(\rho_2,\sigma_2)}^2\,,
\end{equation}
and the claim follows.
\end{proof}
\end{prop}
\begin{thm}\label{thm:zero}
For any $\rho,\,\sigma\in\mathcal{S}(\mathcal{H})$ with finite energy and any $\Pi\in\mathcal{C}(\rho,\sigma)$,
\begin{equation}
C(\Pi) \ge \frac{1}{2}\left(C\left(\left\|\sqrt{\rho}\right\rrangle\left\llangle\sqrt{\rho}\right\|\right) + C\left(\left\|\sqrt{\sigma}\right\rrangle\left\llangle\sqrt{\sigma}\right\|\right)\right)\,.
\end{equation}
\begin{proof}
We consider the case $N=1$, the extension to generic $N$ being straightforward.
Let
\begin{equation}
\Pi = \sum_{n=0}^\infty p_n\,\left\|X_n\right\rrangle\left\llangle X_n\right\|
\end{equation}
be an eigendecomposition of $\Pi$ as a linear operator on the Hilbert space $\mathcal{T}_2(\mathcal{H})$.
Since $\rho$ and $\sigma$ have finite energy, from \autoref{prop:CE} $\Pi$ has finite cost.
We then have
\begin{align}
C(\Pi) &= \mathrm{Tr}_{\mathcal{H}\otimes\mathcal{H}^*}\left[\left(R\otimes\mathbb{I}_{\mathcal{H}^*} - \mathbb{I}_{\mathcal{H}}\otimes R^T\right)\Pi\left(R\otimes\mathbb{I}_{\mathcal{H}^*} - \mathbb{I}_{\mathcal{H}}\otimes R^T\right)\right]\nonumber\\
&= \sum_{n=0}^\infty p_n\left\|\left(R\otimes\mathbb{I}_{\mathcal{H}^*} - \mathbb{I}_{\mathcal{H}}\otimes R^T\right)\left\|X_n\right\rrangle\right\|^2 \overset{\textnormal{(a)}}{=} \sum_{n=0}^\infty p_n\left\|\left[R,\,X_n\right]\right\|_2^2\nonumber\\
&\overset{\textnormal{(b)}}{\ge} \sum_{n=0}^\infty p_n\,\mathrm{Tr}_{\mathcal{H}}\left[R\left(X_n^\dag\,X_n + X_n\,X_n^\dag\right)R - \sqrt{X_n^\dag\,X_n}\,R\,\sqrt{X_n^\dag\,X_n}\,R\right.\nonumber\\
&\phantom{\overset{\textnormal{(b)}}{\ge} \sum_{n=0}^\infty p_n\,\mathrm{Tr}_{\mathcal{H}}} \quad \left. - \sqrt{X_n\,X_n^\dag}\,R\,\sqrt{X_n\,X_n^\dag}\,R\right]\nonumber\\
&\overset{\textnormal{(c)}}{=}\mathrm{Tr}_{\mathcal{H}}\left[R\left(\rho+\sigma\right)R\right]\nonumber\\
&\phantom{\overset{\textnormal{(c)}}{=}}\; - \sum_{n=0}^\infty p_n\,\mathrm{Tr}_{\mathcal{H}}\left[\sqrt{X_n^\dag\,X_n}\,R\,\sqrt{X_n^\dag\,X_n}\,R - \sqrt{X_n\,X_n^\dag}\,R\,\sqrt{X_n\,X_n^\dag}\,R\right]\,,
\end{align}
where (a) follows from \autoref{lem:AXB}, (b) follows from \autoref{lem:RX} of \autoref{sec:app} and (c) follows since
\begin{equation}
\rho = \left(\mathrm{Tr}_{\mathcal{H}}\Pi\right)^T = \sum_{n=0}^\infty p_n\,X_n^\dag\,X_n\,,\qquad \sigma = \mathrm{Tr}_{\mathcal{H}^*}\Pi = \sum_{n=0}^\infty p_n\,X_n\,X_n^\dag\,,
\end{equation}
where both series converge in the trace norm.
Lieb's concavity theorem \cite{lieb1973convex} implies
\begin{align}
\sum_{n=0}^\infty p_n\,\mathrm{Tr}_{\mathcal{H}}\left[\sqrt{X_n^\dag\,X_n}\,R\,\sqrt{X_n^\dag\,X_n}\,R\right] &\le \mathrm{Tr}_{\mathcal{H}}\left[\sqrt{\rho}\,R\,\sqrt{\rho}\,R\right]\,,\nonumber\\
\sum_{n=0}^\infty p_n\,\mathrm{Tr}_{\mathcal{H}}\left[\sqrt{X_n\,X_n^\dag}\,R\,\sqrt{X_n\,X_n^\dag}\,R\right] &\le \mathrm{Tr}_{\mathcal{H}}\left[\sqrt{\sigma}\,R\,\sqrt{\sigma}\,R\right]\,,
\end{align}
therefore
\begin{align}
C(\Pi) &\ge \mathrm{Tr}_{\mathcal{H}}\left[R\left(\rho+\sigma\right)R - \sqrt{\rho}\,R\,\sqrt{\rho}\,R - \sqrt{\sigma}\,R\,\sqrt{\sigma}\,R\right]\nonumber\\
&= \frac{1}{2}\left(\left\|\left[R,\,\sqrt{\rho}\right]\right\|_2^2 + \left\|\left[R,\,\sqrt{\sigma}\right]\right\|_2^2\right)\nonumber\\
&= \frac{1}{2}\left(C\left(\left\|\sqrt{\rho}\right\rrangle\left\llangle\sqrt{\rho}\right\|\right) + C\left(\left\|\sqrt{\sigma}\right\rrangle\left\llangle\sqrt{\sigma}\right\|\right)\right)\,,
\end{align}
where we have used \autoref{lem:AXB} again, and the claim follows.
\end{proof}
\end{thm}
The fundamental consequence of \autoref{thm:zero} is that the identity channel is the optimal plan to transport a quantum state on itself.
\begin{cor}[trivial transport]
For any $\rho\in\mathcal{S}(\mathcal{H})$ with finite energy, the optimal plan to transport $\rho$ onto itself is the identity channel and
\begin{equation}
{D(\rho,\rho)}^2 = C\left(\left\|\sqrt{\rho}\right\rrangle\left\llangle\sqrt{\rho}\right\|\right)\,.
\end{equation}
\end{cor}

We can define the composition of quantum transport plans through the composition of the associated quantum channels.
The possibility of composing quantum transport plans allows us to prove the following modified triangle inequality for the quantum Wasserstein distance.
\begin{thm}[modified triangle inequality]\label{thm:triangle}
For any $\rho_A,\,\rho_B,\,\rho_C\in\mathcal{S}(\mathcal{H})$ with finite energy,
\begin{equation}
D(\rho_A,\rho_C)\le D(\rho_A,\rho_B) + D(\rho_B,\rho_B) + D(\rho_B,\rho_C)\,.
\end{equation}
\begin{proof}
It is sufficient to prove the thesis for quantum states with finite dimensional support and obtain the general case as a limit by \autoref{thm:continuity-distance} and \autoref{rem:finite-approx} of \autoref{sec:continuity}.

For the sake of a simpler notation, we consider $\rho_A$, $\rho_B$ and $\rho_C$ as operators on the Hilbert spaces $\mathcal{H}_A$, $\mathcal{H}_B$ and $\mathcal{H}_C$, respectively (each of which is canonically isomorphic to $\mathcal{H}$).
Let $\Phi_{A\to B}\in\mathcal{M}(\rho_A,\rho_B)$ and $\Phi_{B\to C}\in\mathcal{M}(\rho_B,\rho_C)$, such that
\begin{equation}
\Phi_{A\to C} = \Phi_{B\to C}\circ\Phi_{A\to B}\in\mathcal{M}(\rho_A,\rho_C)\,.
\end{equation}
Let also
\begin{align}
\Pi_{BA^*} &= \Pi_{\Phi_{A\to B}}\in\mathcal{C}(\rho_A,\rho_B)\,,\nonumber\\
\Pi_{CB^*} &= \Pi_{\Phi_{B\to C}}\in\mathcal{C}(\rho_B,\rho_C)\,,\nonumber\\
\Pi_{CA^*} &= \Pi_{\Phi_{A\to C}}\in\mathcal{C}(\rho_A,\rho_C)\,.
\end{align}
We have
\begin{align}
\Pi_{CA^*} &= \left(\Phi_{B\to C}\otimes\mathbb{I}_{\mathcal{T}_1(\mathcal{H}_A^*)}\right)(\Pi_{BA^*})= \mathrm{Tr}_{B^*}\left[\left(\mathbb{I}_{\mathcal{H}_C}\otimes\Psi_{BA^*}^{TB}\right)\left(\Pi_{CB^*}\otimes\mathbb{I}_{\mathcal{H}_A^*}\right)\right] \nonumber\\
&= \left\llangle\sqrt{\rho_B}\right\| \left(\Psi_{CB^*}\otimes\Psi_{BA^*}\right)\left\|\sqrt{\rho_B}\right\rrangle\,,
\end{align}
where $TB$ denotes the partial transposition on $\mathcal{H}_B$, $\left\|\sqrt{\rho_B}\right\rrangle\in\mathcal{H}_B\otimes\mathcal{H}_B^*$ and
\begin{align}\label{eq:pseudo-inverse}
\Psi_{CB^*} &= \left(\mathbb{I}_{\mathcal{H}_C}\otimes\rho_B^{-\frac{T}{2}}\right) \Pi_{CB^*} \left(\mathbb{I}_{\mathcal{H}_C}\otimes\rho_B^{-\frac{T}{2}}\right)\,,\nonumber\\
\Psi_{BA^*} &= \left(\rho_B^{-\frac{1}{2}}\otimes \mathbb{I}_{\mathcal{H}_A^*}\right) \Pi_{BA^*} \left(\rho_B^{-\frac{1}{2}}\otimes \mathbb{I}_{\mathcal{H}_A^*}\right)\,,
\end{align}
where $\rho_B^{-1}$ is defined as the Moore–Penrose inverse \cite{moore1920reciprocal} of $\rho_B$, which is identically zero on the orthogonal of the support of $\rho_B$ and is equal to the inverse of $\rho_B$ on its support.
We notice that $\rho_B^{-1}$ is bounded since $\rho_B$ has finite rank.

We consider the case of one quadrature operator $R$, \emph{i.e.}, $N=1$, the extension to generic $N$ being straightforward.
Let $R_A$, $R_B$ and $R_C$ be the operator $R$ acting on $\mathcal{H}_A$, $\mathcal{H}_B$ and $\mathcal{H}_C$, respectively.
We have from the triangle inequality for the Hilbert norm
\begin{align}
D(\rho_A,\rho_C) &\le \sqrt{\mathrm{Tr}_{CA^*}\left[\left(R_C - R_A^T\right)\Pi_{CA^*}\left(R_C - R_A^T\right)\right]}\nonumber\\
&=\sqrt{\mathrm{Tr}_{CA^*}\left[ \left\llangle\sqrt{\rho_B}\right\| \left(R_C - R_A^T\right) \Psi_{CB^*}\,\Psi_{BA^*} \left(R_C - R_A^T\right) \left\|\sqrt{\rho_B}\right\rrangle\right]}\nonumber\\
& = \left\|\sqrt{\Psi_{CB^*}\,\Psi_{BA^*}} \left(R_C - R_A^T\right)\left\|\sqrt{\rho_B}\right\rrangle\right\|\nonumber\\
&\le \left\|\sqrt{\Psi_{CB^*}\,\Psi_{BA^*}} \left(R_C - R_B\right)\left\|\sqrt{\rho_B}\right\rrangle\right\|\nonumber\\
&\phantom{\le} + \left\|\sqrt{\Psi_{CB^*}\,\Psi_{BA^*}} \left(R_B - R_B^T\right)\left\|\sqrt{\rho_B}\right\rrangle\right\|\nonumber\\
&\phantom{\le}+\left\|\sqrt{\Psi_{CB^*}\,\Psi_{BA^*}} \left(R_B^T - R_A^T\right)\left\|\sqrt{\rho_B}\right\rrangle\right\|\nonumber\\
&= \sqrt{\mathrm{Tr}_{CB^*}\left[\left(R_C - R_B^T\right)\Pi_{CB^*}\left(R_C - R_B^T\right)\right]}\nonumber\\
&\phantom{=} + \sqrt{\mathrm{Tr}_{BA^*}\left[\left(R_B - R_A^T\right)\Pi_{BA^*}\left(R_B - R_A^T\right)\right]}\nonumber\\
&\phantom{=} + \| \left(R_B- R_B^T\right) \left\|\sqrt{\rho_B}\right\rrangle \| ,
\end{align}
and the claim follows taking the infimum over $\Pi_{BA^*}\in\mathcal{C}(\rho_A,\rho_B)$ and $\Pi_{CB^*}\in\mathcal{C}(\rho_B,\rho_C)$.
\end{proof}
\end{thm}

\section{Connection with the Wigner-Yanase metric}\label{sec:WY}
Let $\mathcal{H}$ be a Hilbert space with finite dimension.
The Wigner-Yanase metric \cite{gibilisco2003wigner,bengtsson2017geometry} is the Riemannian metric $g$ on the manifold of quantum states of $\mathcal{H}$ given by
\begin{equation}
g_\rho(X,Y) = 4\left\llangle X\right\|\left(\sqrt{\rho}\otimes\mathbb{I}_{\mathcal{H}^*} +  \mathbb{I}_{\mathcal{H}}\otimes{\sqrt{\rho}}^T\right)^{-2} \left\|Y\right\rrangle
\end{equation}
for any $\rho\in\mathcal{S}(\mathcal{H})$ and any tangent vectors $X$ and $Y$ at the point $\rho$.

The following \autoref{prop:DWY} connects the quantum Wasserstein square distance between a quantum state and itself with the Wigner-Yanase square norms of the tangent vectors induced by the commutators with the quadratures.
\begin{prop}[Wasserstein -- Wigner-Yanase connection]\label{prop:DWY}
For any $\rho\in\mathcal{S}(\mathcal{H})$,
\begin{equation}
{D(\rho,\rho)}^2 = \frac{1}{4}\sum_{i=1}^{N}g_\rho\left(\mathrm{i}\left[R_i,\,\rho\right],\,\mathrm{i}\left[R_i,\,\rho\right]\right)\,.
\end{equation}
\begin{proof}
With the help of \autoref{lem:AXB}, we have for any $i=1,\,\ldots,\,N$
\begin{align}
g_\rho\left(\mathrm{i}\left[R_i,\,\rho\right],\,\mathrm{i}\left[R_i,\,\rho\right]\right) &= -4\,\mathrm{Tr}_{\mathcal{H}}\left[\left[R_i,\,\sqrt{\rho}\right]^2\right]\nonumber\\
&= 4\left\llangle\sqrt{\rho}\right\|\left(R_i\otimes\mathbb{I}_{\mathcal{H}^*}-\mathbb{I}_{\mathcal{H}}\otimes R_i^T\right)^2\left\|\sqrt{\rho}\right\rrangle\,,
\end{align}
and the claim follows.
\end{proof}
\end{prop}

\section{The Wasserstein distance for quantum Gaussian systems}\label{sec:QGS}
We now specialize to quantum Gaussian systems, which provide the mathematical model for the electromagnetic radiation in the quantum regime.
Here we will just give a brief introduction to the required formalism.
For a more comprehensive presentation of quantum Gaussian systems and their applications in quantum information, the reader can consult Refs. \cite{braunstein2005quantum,weedbrook2012gaussian,ferraro2005gaussian,serafini2017quantum,holevo2015gaussian,de2018gaussian}.

\subsection{Introduction to quantum Gaussian systems}\label{sec:intro}
The Hilbert space of a quantum Gaussian system is $\mathcal{H}=L^2(\mathbb{R}^m)$, \emph{i.e.}, the Hilbert space of $m$ harmonic oscillators.
Let $Q_1,\,\ldots,\,Q_m$ and $P_1,\,\ldots,\,P_m$ be the position and momentum operators of the $m$ modes, which act on a wavefunction $\psi\in L^2(\mathbb{R}^m)$ as
\begin{equation}
(Q_i\psi)(q) = q_i\,\psi(q)\,,\qquad (P_i\psi)(q) = -\mathrm{i}\,\frac{\partial}{\partial q_i}\psi(q)
\end{equation}
and satisfy the canonical commutation relations
\begin{equation}
\left[Q_i,\,P_j\right] = \mathrm{i}\,\delta_{ij}\,\mathbb{I}_{\mathcal{H}}\,,\qquad\left[Q_i,\,Q_j\right] =\left[P_i,\,P_j\right] = 0\,,\qquad i,\,j=1,\,\ldots,\,m\,.
\end{equation}
It is useful to define the quadratures
\begin{equation}\label{eq:quadr}
R_1 = Q_1\,,\quad R_2 = P_1\,,\quad\ldots\,,\quad R_{2m-1} = Q_m\,,\quad R_{2m} = P_m\,,
\end{equation}
which satisfy the commutation relations
\begin{equation}
\left[R_i,\,R_j\right] = \mathrm{i}\,\Delta_{ij}\,\mathbb{I}_{\mathcal{H}}\,,\qquad i,\,\ldots,\,j=1,\,\ldots,\,2m\,,
\end{equation}
where
\begin{equation}
\Delta = \bigoplus_{k=1}^m\left(
                            \begin{array}{cc}
                              0 & 1 \\
                              -1 & 0 \\
                            \end{array}
                          \right)
\end{equation}
is the symplectic form.
We also define the ladder operators
\begin{equation}
a_i = \frac{Q_i + \mathrm{i}\,P_i}{\sqrt{2}}\,,\qquad i=1,\,\ldots,\,m\,,
\end{equation}
satisfying the commutation relations
\begin{equation}\label{eq:CCRa}
\left[a_i,\,a_j^\dag\right] = \delta_{ij}\,\mathbb{I}_{\mathcal{H}}\,,\qquad \left[a_i,\,a_j\right] =\left[a_i^\dag,\,a_j^\dag\right] = 0\,,\qquad i,\,j=1,\,\ldots,\,m\,.
\end{equation}

The first moments of a quantum state $\rho$ are the expectation values of the quadratures
\begin{equation}
r_i = \mathrm{Tr}\left[R_i\,\rho\right]\,,\qquad i=1,\,\ldots,\,2m\,,
\end{equation}
and its covariance matrix is
\begin{equation}
\sigma_{ij} = \frac{1}{2}\,\mathrm{Tr}\left[\left\{R_i-r_i\,\mathbb{I}_{\mathcal{H}},\;R_j-r_j\,\mathbb{I}_{\mathcal{H}}\right\}\rho\right]\,,\qquad i,\,j = 1,\,\ldots,\,2m\,,
\end{equation}
where
\begin{equation}
\{X,\,Y\} = X\,Y + Y\,X
\end{equation}
is the anticommutator.
From the Robertson-Heisenberg uncertainty principle, the covariance matrix of any quantum state satisfies \cite{serafini2017quantum}
\begin{equation}\label{eq:uncert}
\sigma \ge \pm\frac{\mathrm{i}}{2}\,\Delta\,.
\end{equation}

A fundamental class of states of quantum Gaussian systems is the class of quantum Gaussian states.
They are the Gibbs thermal states of quadratic Hamiltonians, and they are the easiest states to prepare in the laboratory.
For this reason, they play a key role in several quantum information protocols, \emph{e.g.}, in protocols for quantum key distribution, quantum teleportation or for communication of classical information \cite{weedbrook2012gaussian,ferraro2005gaussian,serafini2017quantum}.
\begin{defn}[quantum Gaussian state]
A quantum Gaussian state of $\mathcal{H}$ is a quantum state proportional to the exponential of a quadratic polynomial in the quadratures:
\begin{equation}\label{eq:defrhoG}
\rho \propto \exp\left(-\frac{1}{2}\sum_{i,\,j=1}^{2m}\left(R_i-r_i\,\mathbb{I}_{\mathcal{H}}\right)h_{ij}\left(R_j - r_j\,\mathbb{I}_{\mathcal{H}}\right)\right)\,,
\end{equation}
where $r\in\mathbb{R}^{2m}$ and $h$ is a strictly positive $2m\times2m$ real matrix.
A quantum Gaussian state is completely determined by its first moments and its covariance matrix: for any $r\in\mathbb{R}^{2m}$ and any symmetric $2m\times2m$ real matrix $\sigma$ satisfying \eqref{eq:uncert}, there exists a unique quantum Gaussian state with first moments $r$ and covariance matrix $\sigma$.

$\mathcal{H}^*$ is the Hilbert space of the quantum Gaussian system with $m$ modes and quadratures $R_1^T,\,\ldots,\,R_{2m}^T$, which satisfy the commutation relations
\begin{equation}\label{eq:CCR*}
\left[R_i^T,\,R_j^T\right] = -\mathrm{i}\,\Delta_{ij}\,\mathbb{I}_{\mathcal{H}^*}\,.
\end{equation}
$\mathcal{H}\otimes\mathcal{H}^*$ is the Hilbert space of a quantum Gaussian system with $2m$ modes, and a quantum Gaussian state of $\mathcal{H}\otimes\mathcal{H}^*$ is a quantum state proportional to the exponential of a quadratic polynomial in $R_i$ and $R_i^T$.
\end{defn}
A special class of quantum Gaussian states are the thermal quantum Gaussian states, for which both the covariance matrix and the matrix $h$ in \eqref{eq:defrhoG} are proportional to the identity.
The thermal quantum Gaussian state with zero temperature is the vacuum state $|0\rangle\langle0|$, which is the projector onto the ground state $|0\rangle$ of the photon-number Hamiltonian
\begin{equation}
H = \frac{1}{2}\sum_{i=1}^{2m}\left(R_i^2-\frac{\mathbb{I}_{\mathcal{H}}}{2}\right)\,;
\end{equation}
its covariance matrix is $\frac{1}{2}I_{2m}$.

The quantum Gaussian unitary operators are the unitary operators that preserve the set of quantum Gaussian states.
The main quantum Gaussian unitary operators are the displacement operators, the beamsplitter and the squeezing.
For any $z\in\mathbb{C}^m$, the displacement operator
\begin{equation}\label{eq:D}
D(z)=\exp\left(\sum_{i=1}^m\left(z_i\,a_i^\dag - z_i^*\,a_i\right)\right)
\end{equation}
is the unitary operator that acts on the ladder operators as \cite{barnett2002methods}
\begin{equation}\label{eq:defD}
{D(z)}^\dag\,a_i\,D(z) = a_i + z_i\,\mathbb{I}_{\mathcal{H}}\,.
\end{equation}
The beamsplitter and the squeezing are the quantum counterparts of the classical linear mixing of random variables, and are the main transformations in quantum optics.
Let $A$ and $B$ be $m$-mode quantum Gaussian systems with Hilbert spaces $\mathcal{H}_A$ and $\mathcal{H}_B$ and ladder operators $a_1\ldots a_m$ and $b_1\ldots b_m$, respectively.
The \emph{beamsplitter} of transmissivity $0\le\eta\le1$ is implemented by the unitary operator
\begin{equation}\label{eq:defU}
U(\eta)=\exp\left(\arccos\sqrt{\eta}\sum_{i=1}^m\left(a_i^\dag b_i-b_i^\dag a_i\right)\right)\,,
\end{equation}
and performs a linear rotation of the ladder operators \cite[Section 1.4.2]{ferraro2005gaussian}:
\begin{align}\label{eq:defUlambda}
{U(\eta)}^\dag\,a_i\,U(\eta) &= \sqrt{\eta}\,a_i+\sqrt{1-\eta}\,b_i\,,\nonumber\\
{U(\eta)}^\dag\,b_i\,U(\eta) &= -\sqrt{1-\eta}\,a_i+\sqrt{\eta}\,b_i\,,\qquad i=1,\,\ldots,\,m\,.
\end{align}
We define for any $0\le\eta\le1$ and any $\rho\in\mathcal{S}\left(\mathcal{H}_A\otimes\mathcal{H}_B\right)$
\begin{equation}\label{eq:defB}
\mathcal{B}_\eta(\rho) = \mathrm{Tr}_B\left[U(\eta)\,\rho\,{U(\eta)}^\dag\right]\,.
\end{equation}
The \emph{squeezing} \cite{barnett2002methods} of parameter $\kappa\ge1$ is implemented by the unitary operator
\begin{equation}\label{eq:defUk}
U(\kappa)=\exp\left(\mathrm{arccosh}\sqrt{\kappa}\sum_{i=1}^m\left(a_i^\dag b_i^\dag-a_i\,b_i\right)\right)\,,
\end{equation}
and acts on the ladder operators as
\begin{align}
{U(\kappa)}^\dag\,a_i\,U(\kappa) &= \sqrt{\kappa}\,a_i+\sqrt{\kappa-1}\,b_i^\dag\,,\nonumber\\
{U(\kappa)}^\dag\,b_i\,U(\kappa) &= \sqrt{\kappa-1}\,a_i^\dag+\sqrt{\kappa}\,b_i\,,\qquad i=1,\,\ldots,\,m\,.
\end{align}

Quantum Gaussian channels are the quantum channels that preserve the set of quantum Gaussian states, and provide the mathematical model for the attenuation and the noise that affect electromagnetic signals traveling through optical fibers and for their amplification.
The most important families of quantum Gaussian channels are the quantum Gaussian attenuators and amplifiers.
The \emph{noiseless quantum Gaussian attenuator} $\mathcal{E}_{\eta}$ \cite[case (C) with $k=\sqrt{\eta}$ and $N=0$]{holevo2007one} models the attenuation affecting electromagnetic signals traveling through optical fibers or free space and can be implemented mixing the input state $\rho$ with the vacuum state through a beamsplitter of transmissivity $0\le\eta\le1$:
\begin{equation}\label{eq:defE}
\mathcal{E}_{\eta}(\rho) = \mathcal{B}_\eta(\rho\otimes|0\rangle\langle0|)\,.
\end{equation}
The \emph{noiseless quantum Gaussian amplifier} $\mathcal{A}_{\kappa}$ \cite[case (C) with $k=\sqrt{\kappa}$ and $N=0$]{holevo2007one} models the amplification of electromagnetic signals and can be implemented performing a squeezing of parameter $\kappa\ge1$ on the input state $\rho$ and the vacuum state:
\begin{equation}\label{eq:defA}
\mathcal{A}_{\kappa}(\rho) = \mathrm{Tr}_{\mathcal{H}_B}\left[U(\kappa)\left(\rho\otimes|0\rangle\langle0|\right){U(\kappa)}^\dag\right]\,.
\end{equation}

\subsection{The Wasserstein distance}\label{sec:WG}
We are now ready to define the Wasserstein distance for quantum Gaussian systems.
In analogy to the classical transport cost on $\mathbb{R}^n$, we consider the cost associated to the canonical quadratures \eqref{eq:quadr}, such that the cost operator is
\begin{equation}\label{eq:CG}
C = \frac{1}{2}\sum_{i=1}^{2m}\left(R_i\otimes\mathbb{I}_{\mathcal{H}^*} - \mathbb{I}_{\mathcal{H}}\otimes R_i^T\right)^2\,.
\end{equation}
\eqref{eq:CG} differs from \eqref{eq:C} by a factor $\frac{1}{2}$ to match the classical normalization and the GMPC normalization.
Sometimes it will be useful to express the cost operator in terms of the ladder operators:
\begin{equation}
C = \sum_{i=1}^{m}\left(a_i\otimes\mathbb{I}_{\mathcal{H}^*} - \mathbb{I}_{\mathcal{H}}\otimes a_i^T\right)^\dag\left(a_i\otimes\mathbb{I}_{\mathcal{H}^*} - \mathbb{I}_{\mathcal{H}}\otimes a_i^T\right)\,.
\end{equation}

Contrarily to the cost operator adopted by GMPC, our $C$ does not have discrete eigenvalues, and its essential spectrum is the whole interval $[0,\infty)$.
As in the classical case, since $C$ is a quadratic polynomial in the quadratures, the cost of a quantum coupling is completely determined by its first moments and its covariance matrix.
Therefore, as for the transport distance between classical Gaussian probability measures, the transport distance between quantum Gaussian states can be computed considering only Gaussian couplings:
\begin{prop}\label{prop:G}
Let $\rho$ and $\sigma$ be quantum Gaussian states of $\mathcal{H}$.
Then, the infimum over $\mathcal{C}(\rho,\sigma)$ in the definition of $D(\rho,\sigma)$ can be restricted to the quantum Gaussian states of $\mathcal{H}\otimes\mathcal{H}^*$ with marginals $\sigma$ and $\rho^T$, respectively.
\begin{proof}
If we replace a generic $\Pi\in\mathcal{C}(\rho,\sigma)$ with the quantum Gaussian state with the same first moments and covariance matrix, both the marginals and the cost remain the same.
\end{proof}
\end{prop}

The optimization over quantum Gaussian couplings can be performed analytically when $\rho$ and $\sigma$ are thermal quantum Gaussian states, and the optimal transport plans are noiseless quantum Gaussian attenuators or amplifiers:
\begin{thm}\label{prop:thermal}
For any $\nu\ge\frac{1}{2}$, let $\omega(\nu)$ be the thermal quantum Gaussian state with covariance matrix $\nu\,I_{2m}$.
Then, for any $\frac{1}{2}\le \nu\le \nu'$,
\begin{itemize}
\item the optimal transport plan from $\omega(\nu)$ to $\omega(\nu')$ is the noiseless quantum Gaussian amplifier $\mathcal{A}_{\kappa}$ with amplification parameter $\kappa=\frac{2\nu'+1}{2\nu+1}$;
\item the optimal transport plan from $\omega(\nu)$ to $\omega(\nu')$ is the noiseless quantum Gaussian attenuator $\mathcal{E}_\eta$ with attenuation parameter $\eta=\frac{2\nu-1}{2\nu'-1}$;
\item in both cases, the transport distance is
\begin{equation}\label{eq:DG}
D(\omega(\nu),\,\omega(\nu')) = \sqrt{m}\left(\sqrt{\nu'+\tfrac{1}{2}} - \sqrt{\nu-\tfrac{1}{2}}\right)\,.
\end{equation}
\end{itemize}
\begin{proof}
Thanks to \autoref{prop:additivity}, it is sufficient to prove the claim for $m=1$.
Thanks to \autoref{prop:G}, we can assume that the coupling $\Pi$ is a quantum Gaussian state.
Its covariance matrix $\sigma$ must have the form
\begin{equation}
\sigma(X) = \left(
           \begin{array}{cc}
             \nu'\,I_2 & X \\
             X^T & \nu\,I_2 \\
           \end{array}
         \right)\,,
\end{equation}
where $X$ is a $2\times2$ real matrix such that
\begin{equation}\label{eq:sigmaX}
\sigma(X)\ge\pm\frac{\mathrm{i}}{2}\left(
                                     \begin{array}{cc}
                                       \Delta & 0 \\
                                       0 & -\Delta \\
                                     \end{array}
                                   \right)
\,,
\end{equation}
where the minus sign on the symplectic matrix of $\mathcal{H}^*$ is due to the commutation relations \eqref{eq:CCR*}.
The cost associated to the coupling above is
\begin{equation}
\mathrm{Tr}_{\mathcal{H}\otimes\mathcal{H}^*}\left[C\,\Pi(X)\right] = \nu' + \nu - \mathrm{tr}\,X\,.
\end{equation}
For any $X$ satisfying \eqref{eq:sigmaX}, $X'=\frac{\mathrm{tr}X}{2}\,I_2$ still satisfies \eqref{eq:sigmaX}, and the associated coupling has the same cost.
Therefore, we can assume that $X = c\,I_2$ for some $c\in\mathbb{R}$.
The condition \eqref{eq:sigmaX} becomes $c^2\le \left(\nu-\frac{1}{2}\right)\left(\nu'+\frac{1}{2}\right)$, hence the optimal coupling has covariance matrix
\begin{equation}
\sigma^* = \left(
           \begin{array}{cc}
             \nu'\,I_2 & \sqrt{\left(\nu-\frac{1}{2}\right)\left(\nu'+\frac{1}{2}\right)}\,I_2 \\
             \sqrt{\left(\nu-\frac{1}{2}\right)\left(\nu'+\frac{1}{2}\right)}\,I_2 & \nu\,I_2 \\
           \end{array}
         \right)\,,
\end{equation}
and the claim \eqref{eq:DG} follows.

For any $\nu\ge\frac{1}{2}$, let
\begin{equation}
\gamma(\nu) = \left\|\sqrt{\omega(\nu)}\right\rrangle\left\llangle\sqrt{\omega(\nu)}\right\|\,.
\end{equation}
$\gamma(\nu)$ is the quantum Gaussian state with zero first moments and covariance matrix
\begin{equation}
\sigma(\nu) = \left(
           \begin{array}{cc}
             \nu\,I_2 & \sqrt{\left(\nu-\frac{1}{2}\right)\left(\nu+\frac{1}{2}\right)}\,I_2 \\
             \sqrt{\left(\nu-\frac{1}{2}\right)\left(\nu+\frac{1}{2}\right)}\,I_2 & \nu\,I_2 \\
           \end{array}
         \right)\,.
\end{equation}
From \cite[Corollary 1]{de2019squashed}, the quantum Gaussian states $(\mathcal{A}_\kappa\otimes\mathbb{I}_{\mathcal{H}^*})(\gamma(\nu))$ and $(\mathbb{I}_{\mathcal{H}}\otimes\mathcal{E}_\eta)(\gamma(\nu'))$ for $\kappa=\frac{2\nu'+1}{2\nu+1}$ and $\eta=\frac{2\nu-1}{2\nu'-1}$
both have covariance matrix equal to $\sigma^*$.
Therefore, they coincide with the optimal coupling, and $\mathcal{A}_\kappa$ and $\mathcal{E}_\eta$ are the optimal plans.
\end{proof}
\end{thm}
The following \autoref{thm:ou} states that the quantum Wasserstein distance is convex with respect to the mixing with the beamsplitter.
\begin{thm}[beamsplitter convexity]\label{thm:ou}
Let $\rho_0,\,\rho_1,\,\sigma_0,\,\sigma_1\in\mathcal{S}(\mathcal{H})$ with finite energy such that for any $i=1,\,\ldots,\,m$
\begin{equation}
\mathrm{Tr}\left[\left(\sigma_0-\rho_0\right)R_i\right] = \mathrm{Tr}\left[\left(\sigma_1-\rho_1\right)R_i\right] = 0\,,
\end{equation}
and for any $0\le\eta\le1$ let
\begin{equation}
\rho_\eta = \mathcal{B}_\eta(\rho_1\otimes\rho_0)\,,\qquad \sigma_\eta = \mathcal{B}_\eta(\sigma_1\otimes\sigma_0)\,,
\end{equation}
where $\mathcal{B}_\eta$ is the beamsplitter \eqref{eq:defB}.
Then,
\begin{equation}
{D(\rho_\eta,\sigma_\eta)}^2 \le \eta\,{D(\rho_1,\sigma_1)}^2 + \left(1-\eta\right){D(\rho_0,\sigma_0)}^2\,.
\end{equation}
\begin{proof}
For any $\Pi_0\in\mathcal{C}(\rho_0,\sigma_0)$ and any $\Pi_1\in\mathcal{C}(\rho_1,\sigma_1)$, let
\begin{equation}
\Pi_\eta = \mathrm{Tr}_{{\mathcal{H}_B}\otimes {\mathcal{H}_B^*}}\left[\left(U(\eta)\otimes U(\eta)^{T\dag}\right)\left(\Pi_1\otimes\Pi_0\right)\left({U(\eta)}^\dag\otimes U(\eta)^T\right)\right]\,.
\end{equation}
Let $C_A$ and $C_B$ be the cost operators of the quantum Gaussian systems $A$ and $B$, respectively.
We have
\begin{align}
&\left({U(\eta)}^\dag\otimes U(\eta)^T\right)C_A\left(U(\eta)\otimes U(\eta)^{T\dag}\right)\nonumber\\
&= \eta\,C_A + \left(1-\eta\right)C_B + \sqrt{\eta\left(1-\eta\right)}\sum_{i=1}^{2m}\left(R_i^A - R_i^{AT}\right)\otimes\left(R_i^B - R_i^{BT}\right)\,.
\end{align}
We have $\Pi_\eta\in\mathcal{C}(\rho_\eta,\sigma_\eta)$, hence
\begin{align}
&{D(\rho_\eta,\sigma_\eta)}^2 \le C(\Pi_\eta) = \eta\,C(\Pi_1) + \left(1-\eta\right)C(\Pi_0)\nonumber\\
&+ \sqrt{\eta\left(1-\eta\right)}\sum_{i=1}^{2m}\mathrm{Tr}_{\mathcal{H}_A\otimes\mathcal{H}_A^*}\left[\left(\sigma_1-\rho_1\right)R_i^A\right] \mathrm{Tr}_{\mathcal{H}_B\otimes\mathcal{H}_B^*}\left[\left(\sigma_0-\rho_0\right)R_i^B\right]\,,
\end{align}
and the claim follows.
\end{proof}
\end{thm}

The following \autoref{thm:heat} states that the quantum Wasserstein distance is subadditive with respect to the addition of classical noise.
\begin{thm}[classical noise]\label{thm:heat}
Let $\rho_0,\,\sigma_0\in\mathcal{S}(\mathcal{H})$ with finite energy and let $\mu,\,\nu$ be probability measures on $\mathbb{C}^{m}$ such that
\begin{align}
&\left(\mathbb{E}_{Z\sim\mu}Z - \mathbb{E}_{W\sim\nu}W\right)\mathrm{Tr}\left[\left(\rho_0-\sigma_0\right)R_i\right] = 0\,,\nonumber\\
&\mathbb{E}_{Z\sim\mu}\|Z\|^2 <\infty\,,\quad \mathbb{E}_{W\sim\nu}\|W\|^2<\infty\,.
\end{align}
Let $\rho_1$ and $\sigma_1$ be the quantum states obtained adding classical noise distributed according to $\mu$ and $\nu$  to $\rho_0$ and $\sigma_0$, respectively:
\begin{equation}
\rho_1 = \int_{\mathbb{C}^{m}}D(z)\,\rho_0\,{D(z)}^\dag\,\mathrm{d}\mu(z)\,,\qquad \sigma_1 = \int_{\mathbb{C}^{m}}D(w)\,\sigma_0\,{D(w)}^\dag\,\mathrm{d}\nu(w)\,,
\end{equation}
where $D(z)$ is the displacement operator \eqref{eq:D}.
Then,
\begin{equation}
{D(\rho_1,\sigma_1)}^2 \le {D(\rho_0,\sigma_0)}^2 + {W(\mu,\nu)}^2\,,
\end{equation}
where $W$ denotes the classical Wasserstein distance.
\begin{proof}
Let $\Pi_0\in\mathcal{C}(\rho_0,\sigma_0)$ and $\pi\in\mathcal{C}(\mu,\nu)$, and let
\begin{equation}
\Pi_1 = \int_{\mathbb{C}^m\times\mathbb{C}^m}\left(D(w)\otimes {D(z)}^{\dag T}\right)\Pi_0\left({D(w)}^\dag\otimes {D(z)}^T\right)\mathrm{d}\pi(z,w)\in\mathcal{C}(\rho_1,\sigma_1)\,.
\end{equation}
With the help of \eqref{eq:defD}, we have for any $z,\,w\in\mathbb{C}^m$
\begin{align}
&\left({D(w)}^\dag\otimes {D(z)}^T\right)C\left(D(w)\otimes {D(z)}^{\dag T}\right)\nonumber\\
&= C + \sum_{i=1}^m\left(\left(w_i-z_i\right)^*\left(a_i-a_i^T\right) + \left(w_i-z_i\right)\left(a_i-a_i^T\right)^\dag\right) + \left|w-z\right|^2\,,
\end{align}
therefore
\begin{equation}
C(\Pi_1) = C(\Pi_0) + \mathbb{E}_{(Z,W)\sim\pi}\left|Z-W\right|^2\,,
\end{equation}
and the claim follows taking the infimum over $\Pi_0\in\mathcal{C}(\rho_0,\sigma_0)$ and $\pi\in\mathcal{C}(\mu,\nu)$.
\end{proof}
\end{thm}

\begin{cor}
The quantum Wasserstein distance is translation invariant, \emph{i.e.}, for any $\rho,\,\sigma\in\mathcal{S}(\mathcal{H})$ with finite energy and any $z\in\mathbb{C}^m$,
\begin{equation}
D\left(D(z)\,\rho\,{D(z)}^\dag,\,D(z)\,\sigma\,{D(z)}^\dag\right) = D(\rho,\sigma)\,.
\end{equation}
\begin{proof}
The claim follows from \autoref{thm:heat} with both $Z$ and $W$ equal to $z$ with probability one.
\end{proof}
\end{cor}

\subsection{Semiclassical limit}\label{sec:classical}
The coherent states of a quantum Gaussian system are the pure quantum Gaussian states obtained applying a displacement operator to the vacuum state and are the eigenvectors of the ladder operators:
\begin{equation}\label{eq:defcoh}
|z\rangle = D(z)|0\rangle\,,\qquad a_i|z\rangle = z_i|z\rangle\,,\qquad z\in\mathbb{C}^m\,,\quad i=1,\,\ldots,\,m\,.
\end{equation}
They are the easiest state to realize in laboratory, and they are considered to be the most classical pure states of the system.
Coherent states form an overcomplete set and satisfy the resolution of the identity \cite{holevo2015gaussian}
\begin{equation}\label{eq:cohI}
\int_{\mathbb{C}^m}|z\rangle\langle z|\,\frac{\mathrm{d}z}{\pi^m} = \mathbb{I}_{\mathcal{H}}\,,
\end{equation}
where the integral converges weakly.
Therefore, for any quantum state $\rho$, the function
\begin{equation}
Q(z) = \langle z|\rho|z\rangle\,,\qquad z\in\mathbb{C}^m\,,
\end{equation}
called Husimi Q representation of $\rho$ \cite{barnett2002methods,serafini2017quantum}, defines a probability density on $\mathbb{C}^m$ with normalization
\begin{equation}
\int_{\mathbb{C}^m}\langle z|\rho|z\rangle\,\frac{\mathrm{d}z}{\pi^m}=1\,.
\end{equation}
A quantum state is completely determined by its Q representation.
Moreover, the Q representation is the probability distribution of a particular measurement that can be performed on the state, which is called heterodyne measurement \cite{barnett2002methods,serafini2017quantum}, and is one of the main measurements in quantum optics.

A quantum state $\rho$ is called semiclassical if it can be expressed as a convex mixture of coherent states, \emph{i.e.},
\begin{equation}
\rho = \int_{\mathbb{C}^m} |z\rangle\langle z|\,\mathrm{d}\hat{\mu}(z)
\end{equation}
for some probability measure $\hat{\mu}$ on $\mathbb{C}^m$, where the integral converges in the trace norm.
If this is the case, it can be proved that $\hat{\mu}$ is uniquely determined, and is called the Glauber-Sudarshan P representation of $\rho$ \cite{barnett2002methods,serafini2017quantum}.

From their definitions, the Q representation of a semiclassical state is equal to its P representation convolved with the Gaussian function $\mathrm{e}^{-|z|^2}$.

We will prove that as the GMPC distance, our quantum Wasserstein distance is upper bounded by the Wasserstein distance between the P representations and lower bounded by the Wasserstein distance between the Q representations.
In the semiclassical limit the P and Q representations become the same, hence both our distance and the GMPC distance recover the classical Wasserstein distance.

\begin{thm}[P representation]\label{thm:classical1}
Let $\hat{\mu}$ and $\hat{\nu}$ be probability measures on $\mathbb{C}^m$ with $\mathbb{E}_{Z\sim\hat{\mu}}\|Z\|^2<\infty$ and $\mathbb{E}_{W\sim\hat{\nu}}\|W\|^2<\infty$, and let
\begin{equation}
\rho = \int_{\mathbb{C}^m} |z\rangle\langle z|\,\mathrm{d}\hat{\mu}(z)\,,\qquad \sigma = \int_{\mathbb{C}^m} |w\rangle\langle w|\,\mathrm{d}\hat{\nu}(w)
\end{equation}
be the associated semiclassical states, where both integrals converge in the trace norm.
Then,
\begin{equation}
{D(\rho,\sigma)}^2 \le {W(\hat{\mu},\hat{\nu})}^2 + m\,.
\end{equation}
\begin{proof}
We define for any $\hat{\pi}\in\mathcal{C}(\hat{\mu},\hat{\nu})$
\begin{equation}
\Pi = \int_{\mathbb{C}^m\times\mathbb{C}^m} |w\rangle\langle w|\otimes{|z\rangle\langle z|}^T\,\mathrm{d}\hat{\pi}(z,w)\in\mathcal{C}(\rho,\sigma)\,,
\end{equation}
where the integral converges in the trace norm.
With the help of \eqref{eq:defcoh}, we have for any $z,\,w\in\mathbb{C}^m$
\begin{equation}
C\left(|w\rangle\langle w|\otimes{|z\rangle\langle z|}^T\right) = m + \left|z-w\right|^2\,,
\end{equation}
therefore we get from \autoref{lem:invariance-energy}
\begin{equation}
C(\Pi) = m + \int_{\mathbb{C}^m\times\mathbb{C}^m}|w-z|^2\,\mathrm{d}\hat{\pi}(z,w)\,,
\end{equation}
and the claim follows taking the $\inf$ over $\hat{\pi}\in\mathcal{C}(\hat{\mu},\hat{\nu})$.
\end{proof}
\end{thm}
\begin{cor}\label{cor:Prho}
Let $\mu$ be a probability measure on $\mathbb{C}^m$ with $\mathbb{E}_{Z\sim\mu}\|Z\|^2<\infty$, and let
\begin{equation}
\rho = \int_{\mathbb{C}^m} |z\rangle\langle z|\,\mathrm{d}\mu(z)
\end{equation}
be a semiclassical state.
Then,
\begin{equation}
{D(\rho,\rho)}^2 \le m\,.
\end{equation}
\end{cor}

\begin{rem}
In the same hypotheses of \autoref{cor:Prho}, \cite[Theorem 2.3]{golse2016mean} implies that
\begin{equation}
D_{\textnormal{GMPC}}(\rho,\rho) = \sqrt{m}\,.
\end{equation}
On the contrary, our distance does not always saturate the upper bound of \autoref{cor:Prho}.
Indeed, if $\rho$ is the $m$-mode quantum thermal Gaussian state $\omega(\nu)$ with covariance matrix $\nu\,I_{2m}$, we get from \autoref{prop:thermal}
\begin{equation}
D(\omega(\nu),\,\omega(\nu)) = \sqrt{m}\left(\sqrt{\nu+\tfrac{1}{2}} - \sqrt{\nu-\tfrac{1}{2}}\right)\,,
\end{equation}
which is strictly decreasing in $\nu$, is equal to $\sqrt{m}$ for $\nu=\frac{1}{2}$ and tends to $0$ for $\nu\to\infty$.
\end{rem}

\begin{thm}[Q representation]\label{thm:classical2}
Let $\rho,\,\sigma\in\mathcal{S}(\mathcal{H})$ with finite energy, and let $\mu$ and $\nu$ be the probability measures on $\mathbb{C}^m$ associated to their respective Husimi Q representations:
\begin{equation}
\mathrm{d}\mu(z) = \langle z|\rho|z\rangle\,\frac{\mathrm{d}z}{\pi^{m}}\,,\qquad \mathrm{d}\nu(w) = \langle w|\sigma|w\rangle\,\frac{\mathrm{d}w}{\pi^{m}}\,,\qquad z,\,w\in\mathbb{C}^m\,.
\end{equation}
Then,
\begin{equation}\label{eq:lbW}
{D(\rho,\sigma)}^2 \ge {W(\mu,\nu)}^2 - m\,.
\end{equation}
\begin{proof}
Let $\pi$ be the probability measure on $\mathbb{C}^m\times\mathbb{C}^m$ given by
\begin{equation}
\mathrm{d}\pi(z,w) = \mathrm{Tr}_{\mathcal{H}\otimes\mathcal{H}^*}\left[\Pi\left(|w\rangle\langle w|\otimes{|z\rangle\langle z|}^T\right)\right]\frac{\mathrm{d}z\,\mathrm{d}w}{\pi^{2m}}\,.
\end{equation}
The marginals of $\pi$ are $\mu$ and $\nu$, therefore $\pi\in\mathcal{C}(\mu,\nu)$.
We have
\begin{align}
&{W(\mu,\nu)}^2 \le \int_{\mathbb{C}^m\times\mathbb{C}^m}\left|w-z\right|^2\,\mathrm{d}\pi(z,w)\nonumber\\
&\overset{(\mathrm{a})}{=} \sum_{i=1}^m\int_{\mathbb{C}^m\times\mathbb{C}^m}\mathrm{Tr}_{\mathcal{H}\otimes\mathcal{H}^*}\left[\left(a_i^\dag\,\Pi\,a_i - a_i^\dag\,a_i^T\,\Pi - \Pi\,a_i\,a_i^{\dag T} + a_i^T\,\Pi\,a_i^{\dag T}\right)\right.\nonumber\\
&\phantom{=\sum_{i=1}^m\int_{\mathbb{C}^m\times\mathbb{C}^m}\mathrm{Tr}_{\mathcal{H}\otimes\mathcal{H}^*}}\quad\left.\left(|w\rangle\langle w|\otimes{|z\rangle\langle z|}^T\right)\right]\frac{\mathrm{d}z\,\mathrm{d}w}{\pi^{2m}}\nonumber\\
&\overset{(\mathrm{b})}{=} \sum_{i=1}^m\mathrm{Tr}_{\mathcal{H}\otimes\mathcal{H}^*}\left[a_i^\dag\,\Pi\,a_i - a_i^\dag\,a_i^T\,\Pi - \Pi\,a_i\,a_i^{\dag T} + a_i^T\,\Pi\,a_i^{\dag T}\right]\nonumber\\
&\overset{(\mathrm{c})}{=} m+\sum_{i=1}^m\mathrm{Tr}_{\mathcal{H}\otimes\mathcal{H}^*}\left[a_i^\dag\,\Pi\,a_i - a_i^\dag\,\Pi\,a_i^T - a_i^{\dag T}\,\Pi\,a_i + a_i^{\dag T}\,\Pi\,a_i^{T}\right]\nonumber\\
&= m+\sum_{i=1}^m\mathrm{Tr}_{\mathcal{H}\otimes\mathcal{H}^*}\left[\left(a_i - a_i^T\right)^\dag\Pi\left(a_i-a_i^T\right)\right] = m + C(\Pi)\,,
\end{align}
where (a) follows from \eqref{eq:defcoh}, (b) follows from \eqref{eq:cohI} and (c) follows from \eqref{eq:CCRa}.
The claim follows taking the $\inf$ over $\Pi\in\mathcal{C}(\rho,\sigma)$.
\end{proof}
\end{thm}

\subsection{Quantum Stam inequality}
The Wasserstein distance between a quantum state and itself satisfies the following Stam inequality.
\begin{thm}[quantum Stam inequality]\label{thm:Stam}
Let $\rho_0,\,\rho_1\in\mathcal{S}(\mathcal{H})$ with finite energy, and for any $0\le\eta\le1$ let
\begin{equation}
\rho_\eta = \mathcal{B}_\eta(\rho_1\otimes\rho_0)\,,
\end{equation}
where $\mathcal{B}_\eta$ is the beamsplitter \eqref{eq:defB}.
Then,
\begin{equation}
\frac{1}{{D(\rho_\eta,\rho_\eta)}^2} \ge \frac{\eta}{{D(\rho_1,\rho_1)}^2} + \frac{1-\eta}{{D(\rho_0,\rho_0)}^2}\,.
\end{equation}
\begin{proof}
Let $\rho\in\mathcal{S}(\mathcal{H})$ have finite energy.
From \autoref{prop:CE} and \autoref{rem:I}, also $\left\|\sqrt{\rho}\right\rrangle\left\llangle\sqrt{\rho}\right\|$ has finite energy, \emph{i.e.}, $\left[R_i,\,\sqrt{\rho}\right]\in\mathcal{T}_2(\mathcal{H})$ for any $i=1,\,\ldots,\,2m$.
Therefore, we can define the Fisher information matrix of $\rho$ as
\begin{equation}
J(\rho)_{ab} = \left\llangle\left[R_a,\,\sqrt{\rho}\right]\left\|\left[R_b,\,\sqrt{\rho}\right]\right.\right\rrangle\,, \qquad a,\,b=1,\,\ldots,\,2m\,,
\end{equation}
such that
\begin{equation}\label{eq:DJ}
{D(\rho,\rho)}^2 = \mathrm{tr}\,J(\rho)\,.
\end{equation}
For any $k\in\mathbb{R}^{2m}$, let
\begin{equation}
\rho(k) = \mathrm{e}^{-\mathrm{i}\sum_{i=1}^{2m}k_i\,R_i}\,\rho\,\mathrm{e}^{\mathrm{i}\sum_{i=1}^{2m}k_i\,R_i}\,.
\end{equation}
We have for any $k,\,q\in\mathbb{R}^{2m}$
\begin{equation}
\mathrm{e}^{\mathrm{i}\sum_{i=1}^{2m}k_i\,R_i}\,\mathrm{e}^{-\mathrm{i}\sum_{i=1}^{2m}q_i\,R_i} \propto \mathrm{e}^{\mathrm{i}\sum_{i=1}^{2m}\left(k_i-q_i\right)R_i}\,,
\end{equation}
therefore, since the Hilbert-Schmidt scalar product is unitarily invariant, we get
\begin{equation}
\left\llangle\sqrt{\rho(k)}\left\|\sqrt{\rho(q)}\right.\right\rrangle = \left\llangle\sqrt{\rho}\left\|\sqrt{\rho(q-k)}\right.\right\rrangle\,,
\end{equation}
and
\begin{align}
J(\rho)_{ab} &= \left.\frac{\partial^2}{\partial k_a\partial q_b}\left\llangle\sqrt{\rho(k)}\left\| \sqrt{\rho(q)}\right.\right\rrangle\right|_{k=q=0}\nonumber\\
&= \left.\frac{\partial^2}{\partial k_a\partial q_b}\left\llangle\sqrt{\rho}\left\| \sqrt{\rho(q-k)}\right.\right\rrangle\right|_{k=q=0}\nonumber\\
&= \left.\frac{\partial^2}{\partial k_a\partial k_b}\left\llangle\sqrt{\rho}\left\|\sqrt{\rho(k)}\right.\right\rrangle\right|_{k=0}\,.
\end{align}
From Lieb's concavity theorem \cite{lieb1973convex}, the function $(\rho,\sigma)\mapsto -\left\llangle\sqrt{\rho}\left\|\sqrt{\sigma}\right.\right\rrangle$ is jointly convex, and therefore contractive with respect to the joint application of a quantum channel to both arguments.
Applying this property to the beamsplitter we get that for any $k,\,q\in\mathbb{R}^{2m}$
\begin{align}\label{eq:sqrt}
&-\left\llangle\sqrt{\rho_1\otimes\rho_0}\left\|\sqrt{\rho_1(k)\otimes\rho_0(q)}\right.\right\rrangle \nonumber\\
&\ge -\left\llangle\sqrt{\mathcal{B}_\eta(\rho_1\otimes\rho_0)}\left\|\sqrt{\mathcal{B}_\eta(\rho_1(k)\otimes\rho_0(q))}\right.\right\rrangle\nonumber\\
&= -\left\llangle\sqrt{\rho_\eta}\left\|\sqrt{\rho_\eta\left(\sqrt{\eta}\,k+\sqrt{1-\eta}\,q\right)}\right.\right\rrangle\,,
\end{align}
where the last equality follows from \cite[Lemma 1]{de2018conditional}, stating that
\begin{equation}
\mathcal{B}_\eta(\rho_1(k)\otimes\rho_0(q)) = \left(\mathcal{B}_\eta(\rho_1\otimes\rho_0)\right)\left(\sqrt{\eta}\,k+\sqrt{1-\eta}\,q\right)\,.
\end{equation}
Since both sides of \eqref{eq:sqrt} are equal to $-1$ for $k=q=0$, which is their global minimum, the inequality \eqref{eq:sqrt} translates to the Hessian with respect to $(k,q)$:
\begin{equation}\label{eq:matrJ}
\left(
  \begin{array}{cc}
    \mathrm{tr}\,J(\rho_1) & 0 \\
    0 & \mathrm{tr}\,J(\rho_0) \\
  \end{array}
\right) \ge \left(
              \begin{array}{c}
                \sqrt{\eta} \\
                \sqrt{1-\eta} \\
              \end{array}
            \right) \mathrm{tr}\,J(\rho_\eta) \left(
                                      \begin{array}{cc}
                                        \sqrt{\eta} & \sqrt{1-\eta} \\
                                      \end{array}
                                    \right)\,.
\end{equation}
Putting together \eqref{eq:matrJ} and \eqref{eq:DJ} we get
\begin{equation}
\left(
  \begin{array}{cc}
    {D(\rho_1,\rho_1)}^2 & 0 \\
    0 & {D(\rho_0,\rho_0)}^2 \\
  \end{array}
\right) \ge \left(
              \begin{array}{c}
                \sqrt{\eta} \\
                \sqrt{1-\eta} \\
              \end{array}
            \right) {D(\rho_\eta,\rho_\eta)}^2 \left(
                                      \begin{array}{cc}
                                        \sqrt{\eta} & \sqrt{1-\eta} \\
                                      \end{array}
                                    \right)\,,
\end{equation}
therefore
\begin{equation}\label{eq:det}
\left(
  \begin{array}{cc}
    {D(\rho_1,\rho_1)}^2 - \eta\,{D(\rho_\eta,\rho_\eta)}^2 & -\sqrt{\eta\left(1-\eta\right)}\,{D(\rho_\eta,\rho_\eta)}^2 \\
    -\sqrt{\eta\left(1-\eta\right)}\,{D(\rho_\eta,\rho_\eta)}^2 & {D(\rho_0,\rho_0)}^2 - \left(1-\eta\right){D(\rho_\eta,\rho_\eta)}^2 \\
  \end{array}
\right) \ge 0\,,
\end{equation}
and the claim follows from the positivity of the determinant of \eqref{eq:det}.
\end{proof}
\end{thm}

\section{Conclusions and perspectives}\label{sec:concl}
We have proposed a new quantum generalization of the Wasserstein distance that has the property that the transport plans are in one-to-one correspondence with quantum channels.
This property allows for the first time to interpret the quantum transport plans as physical operations performed on the system.
We have started to explore the properties of our distance, proving \emph{e.g.} that it satisfies a modified triangle inequality and determining the optimal transport plans between thermal quantum Gaussian states.

The most natural application of our distance is the theory of quantum rate-distortion coding \cite{howard2000ratedistortion,devetak2001quantum,devetak2002quantum,chen2008ratedistortion,datta2013ratedistortion,datta2013classical,wilde2013quantumr,salek2018quantum}, whose goal is to determine the maximum achievable rates for the lossy compression of quantum states with a given distortion.
The distortion measure is defined through the quantum state obtained applying in sequence the encoding and the decoding quantum channel to half of a purification of the source state.
Therefore, the problem of determining the maximum achievable compression rates for a given distortion and a given source state can be related to a sequence of optimal transport problems where the transport plan is given by the composition of the encoding and decoding quantum channel and optimization is performed over both the transport plan and the target state.
So far, most of the effort has focused on the entanglement fidelity as quantum distortion measure.
By contrast, the most common distortion measure in classical rate-distortion theory for signals with values in $\mathbb{R}^n$ is the average square norm of the difference between original and distorted signal \cite[Chapter 10]{cover2006elements}, and coincides with the definition of the transport cost \eqref{eq:defCpi} of a classical coupling.
Our quantum transport cost provides the most natural generalization of the classical distortion measure, and we will explore in future works its applications in this direction.

A further natural application of our distance is the field of quantum machine learning, and in particular the quantum version of the Generative Adversarial Networks (GANs) \cite{lloyd2018quantum}.
Quantum GANs provide an algorithm to learn a target quantum state with a parametric quantum circuit.
A key role is played by the cost function that quantifies the quality of the approximation, \emph{i.e.}, the distance between the target and the generated state.
The distances induced by optimal mass transport have turned out to be the best choice for the cost function of classical GANs \cite{arjovsky2017wasserstein}, and are showing extremely promising results also in the quantum setting \cite{chakrabarti2019quantum,de2020quantum,kiani2021quantum}.
The formulation of the Wasserstein distance with quantum couplings is fundamental to the application to classical GANs, since it allows for a fast algorithm to compute a regularized version of the distance \cite{cuturi2019computational}.
The quantum Wasserstein distance proposed in this paper together with the GMPC distance are defined through a quantum counterpart of the couplings, and therefore constitute very promising candidates for the cost function of quantum GANs.

\subsection*{Acknowledgements}
GdP was supported by the USA Air Force Office of Scientific Research and the USA Army Research Office under the Blue Sky program. DT is a member of Gnampa group (INdAM).
DT has been partially supported by the University of Pisa, Project PRA 2018-49 and Gnampa project 2019 ``Propriet\`a analitiche e geometriche di campi aleatori''.

We thank Luigi Ambrosio for useful remarks on a preliminary version of this work and Emanuele Caglioti for fruitful discussions and for a careful reading of the paper.

\appendix

\section{Continuity of the quantum Wasserstein distance}\label{sec:continuity}

\subsection{Compactness of quantum plans}

Recall that a sequence $\{ A_n \}_{n=0}^\infty \subset \mathcal{B}(\mathcal{H})$ weakly converges  towards $A \in \mathcal{B}(\mathcal{H})$ if, for every $|\phi\rangle, |\psi\rangle \in \mathcal{H}$, \begin{equation}
\lim_{n \to \infty} \langle \phi | A_n| \psi\rangle = \langle \phi | A | \psi\rangle.
\end{equation}

\begin{lem}\label{prop:compactness}
Given $\{\rho_n\}_{n =0 }^\infty \subseteq \mathcal{S}(\mathcal{H})$, there exists $\rho \in \mathcal{T}_1(\mathcal{H})$, $\rho \ge 0$, $\mathrm{Tr}_{\mathcal{H}}\left[\rho\right]\le 1$ and a subsequence $\{\rho_{n_k}\}_{k=0}^\infty$ weakly converging towards $\rho$ . If $\mathrm{Tr}_{\mathcal{H}}\left[\rho\right]=1$ then convergence holds in the trace norm.
\begin{proof}
The first statement is a consequence of Banach-Alaoglu theorem and lower semicontinuity of the trace, see also the proof of \cite[Theorem 11.2]{holevo2019quantum}. The second statement is \cite[Lemma 11.1]{holevo2019quantum}.
\end{proof}
\end{lem}

Because of the result above, we may consider weak convergence or equivalently in the trace norm for a sequence of quantum states $\{\rho_n\}_{n =0}^\infty$, provided that the limit $\rho$ is a quantum state. Hence we often omit to specify which type of convergence we consider on quantum states in the results below.

\begin{prop}\label{lem:compactness-plans}
Let $\rho$, $\sigma \in \mathcal{S}(\mathcal{H})$ be quantum states and let $\{\Pi_n\}_{n=0}^\infty \subset \mathcal{S}(\mathcal{H}\otimes \mathcal{H}^*)$ be a sequence of quantum couplings $\Pi_n \in \mathcal{C}(\rho_n, \sigma_n)$ such that $\lim_n \rho_n = \rho$, $\lim_n \sigma_n = \sigma$. Then, there exists a subsequence $\{ \Pi_{n_k} \}_{k=0}^\infty$ converging towards $\Pi \in \mathcal{C}(\rho, \sigma)$.
\begin{proof}
Existence of a weak limit $\Pi  = \lim_{k} \Pi_{n_k} \in \mathcal{T}_1(\mathcal{H})$, $\Pi \ge 0$, follows from \autoref{prop:compactness}. To argue that $\mathrm{Tr}_{\mathcal{H}\otimes\mathcal{H}^*} [\Pi] = 1$, let $P$, $Q \in \mathcal{B}(\mathcal{H})$ be projectors with $P+Q = \mathbb{I}_{\mathcal{H}}$ and $P$ with finite rank.
We have
\begin{align}
\mathrm{Tr}_{\mathcal{H}\otimes\mathcal{H}^*}[ \Pi] &\ge \mathrm{Tr}_{\mathcal{H}\otimes\mathcal{H}^*}[ (P \otimes P^T) \Pi (P \otimes P^T)]\nonumber\\
&= \lim_{n\to\infty} \mathrm{Tr}_{\mathcal{H}\otimes\mathcal{H}^*}[ (P \otimes P^T) \Pi_n (P \otimes P^T)],
\end{align}
where the limit holds because $P\otimes P^T$ has finite rank. Then
\begin{align}
 & \mathrm{Tr}_{\mathcal{H}\otimes\mathcal{H}^*}[ (P \otimes P^T) \Pi_n (P \otimes P^T)] \nonumber \\
 & = \mathrm{Tr}_{\mathcal{H}\otimes\mathcal{H}^*} [\Pi_n] - \mathrm{Tr}_{\mathcal{H}\otimes\mathcal{H}^*}[ (P \otimes Q^T) \Pi_n (P \otimes Q^T)]\nonumber\\
  &\phantom{=} -\mathrm{Tr}_{\mathcal{H}\otimes\mathcal{H}^*}[ (Q \otimes \mathbb{I}_{\mathcal{H}^*}) \Pi_n (Q \otimes \mathbb{I}_{\mathcal{H}^*})]\nonumber\\
 & = 1 - \mathrm{Tr}_{\mathcal{H}\otimes\mathcal{H}^*}[ (P \otimes Q^T) \Pi_n (P \otimes Q^T)] -\mathrm{Tr}_{\mathcal{H}\otimes\mathcal{H}^*}[ (Q \otimes \mathbb{I}_{\mathcal{H}^*}) \Pi_n (Q \otimes \mathbb{I}_{\mathcal{H}^*})].
\end{align}

We have the inequality
\begin{equation} \mathrm{Tr}_{\mathcal{H}}[(P\otimes Q^T) \Pi_n (P\otimes Q^T)] \le  \mathrm{Tr}_{\mathcal{H}}[(\mathbb{I}_{\mathcal{H}} \otimes Q^T) \Pi_n ( \mathbb{I}_{\mathcal{H}}\otimes Q^T)] \le Q^T \rho_n^T Q^T.
\end{equation}
Taking the partial trace with respect to $\mathcal{H}^*$,
\begin{align}
\limsup_{n\to\infty} \mathrm{Tr}_{\mathcal{H}\otimes\mathcal{H}^*}[ (P \otimes Q^T) \Pi_n (P \otimes Q^T)] &\le \limsup_{n\to\infty} \mathrm{Tr}_{\mathcal{H}^*}[ Q^T \rho^T_n Q^T]\nonumber\\
&= \mathrm{Tr}_{\mathcal{H}^*}[Q^T\rho^TQ^T] = \mathrm{Tr}_{\mathcal{H}}[Q\rho Q].
\end{align}
 Similarly, $\limsup_n \mathrm{Tr}_{\mathcal{H}\otimes\mathcal{H}^*}[ (Q \otimes \mathbb{I}_{\mathcal{H}^*}) \Pi_n (Q \otimes \mathbb{I}_{\mathcal{H}^*})] \le \mathrm{Tr}_{\mathcal{H}}[ Q \sigma Q]$. It follows that
\begin{equation}
\mathrm{Tr}_{\mathcal{H}\otimes\mathcal{H}^*}[ \Pi] \ge 1- \mathrm{Tr}_{\mathcal{H}}[ Q \rho Q]-\mathrm{Tr}_{\mathcal{H}}[ Q \sigma Q].
\end{equation}
Letting $P \to \mathbb{I}_{\mathcal{H}}$ weakly, both $Q\rho Q$ and $Q \sigma Q$ converge to $0$ in the trace norm, hence the thesis.
\end{proof}
\end{prop}

A similar argument gives the following result.

\begin{lem}\label{lem:convergence-subplans}
Given quantum states $\rho$, $\sigma \in \mathcal{S}(\mathcal{H})$ let $\{ P_n \}_{n=0}^\infty$, $\{ Q_n\}_{n=0}^\infty \subset \mathcal{B}(\mathcal{H})$ be such that $0 \le P_n \le \mathbb{I}_{\mathcal{H}}$, $0 \le Q_n \le \mathbb{I}_{\mathcal{H}}$, and $\lim_n P_n\rho P_n = \rho$, $\lim_n Q_n \sigma Q_n= \sigma$.
Then, for any $\Pi \in \mathcal{C}(\rho, \sigma)$,
\begin{equation}\label{eq:convergence-subplans}
\lim_{n\to\infty} \left(Q_n \otimes P^T_n\right) \Pi \left(Q_n \otimes P^T_n\right) = \Pi
\end{equation}
in the trace norm.
\begin{proof}
The assumption $\lim_n P_n \rho P_n = \rho$ implies that $\{P_n\}_{n=0}^\infty$ weakly converges to the identity operator on $\mathrm{supp}\, \rho$. By a density argument, it is sufficient to prove that, for any $|\phi\rangle$ eigenvector for $\rho$, $\rho |\phi\rangle = p | \phi \rangle$, with $p>0$, one has $\lim_n P_n |\phi\rangle = |\phi \rangle$. Since $p |\phi\rangle \langle \phi| \le \rho$, we have indeed
\begin{align}
& \| (\mathbb{I}_{\mathcal{H}} - P_n) |\phi \rangle \|^2 = \mathrm{Tr} [ (\mathbb{I}_{\mathcal{H}} - P_n)  |\phi\rangle \langle \phi|  (\mathbb{I}_{\mathcal{H}} - P_n)] \nonumber \\
&  \le p^{-1} \mathrm{Tr}[(\mathbb{I}_{\mathcal{H}} - P_n) \rho (\mathbb{I}_{\mathcal{H}} - P_n)] \to 0.
\end{align}
Similarly, we have that $\{Q_n\}_{n=0}^\infty$ weakly converges to the identity operator on $\mathrm{supp}\, \sigma$. We deduce that $\{ Q_n \otimes P_n^T \}_{n=0}^\infty$ weakly converges to the identity operator on $\mathrm{supp}\, \sigma \otimes \mathrm{supp}\, \rho^T$, which contains  $\mathrm{supp}\, \Pi$. Hence \eqref{eq:convergence-subplans} holds in the weak sense.

We prove next that
\begin{equation}\label{eq:convergence-trace}
\lim_{n\to\infty} \mathrm{Tr}[ (Q_n \otimes P^T_n) \Pi (Q_n \otimes P^T_n)] = 1.
\end{equation}
We write
\begin{align}
& \mathrm{Tr}[ (Q_n \otimes P^T_n) \Pi (Q_n \otimes P^T_n)] \nonumber\\
&  =  1-\mathrm{Tr}[  \left( Q_n \otimes (\mathbb{I}_{\mathcal{H}^*} - P_n^T)\right) \, \Pi\, \left(  Q_n \otimes ( \mathbb{I}_{\mathcal{H}^*} - P_n^T\right)  ]\nonumber\\
& \phantom{=} -\mathrm{Tr}[ \left(  ( \mathbb{I}_{\mathcal{H}} -Q_n) \otimes  \mathbb{I}_{\mathcal{H}^*} \right) \Pi \left(  (\mathbb{I}_{\mathcal{H}} -Q_n) \otimes \mathbb{I}_{\mathcal{H}^*} \right)].
\end{align}
If we take the partial trace with respect to $\mathcal{H}$, then
\begin{align}
&\mathrm{Tr}_{\mathcal{H}} [  \left( Q_n \otimes (\mathbb{I}_{\mathcal{H}^*} - P_n^T)\right) \, \Pi\, \left(  Q_n \otimes ( \mathbb{I}_{\mathcal{H}^*} - P_n^T\right) ]\nonumber\\
&\le (\mathbb{I}_{\mathcal{H}^*} - P_n^T) \mathrm{Tr}_{\mathcal{H}} [\Pi]  (\mathbb{I}_{\mathcal{H}^*} - P_n^T) = (\rho -P_n \rho P_n)^T,
\end{align}
so that
\begin{equation}
\lim_{n\to\infty} \mathrm{Tr}[  \left( Q_n \otimes (\mathbb{I}_{\mathcal{H}^*} - P_n^T)\right) \, \Pi\, \left(  Q_n \otimes ( \mathbb{I}_{\mathcal{H}^*} - P_n^T\right)  ]  =0.
\end{equation}
Arguing similarly, taking the partial trace with respect to $\mathcal{H}^*$,
\begin{align}
\lim_{n\to\infty} \mathrm{Tr}\left[ \left(  ( \mathbb{I}_{\mathcal{H}} -Q_n) \otimes  \mathbb{I}_{\mathcal{H}^*} \right) \Pi \left(  (\mathbb{I}_{\mathcal{H}} -Q_n) \otimes \mathbb{I}_{\mathcal{H}^*}\right)\right] &\le \lim_{n\to\infty} \mathrm{Tr}_{\mathcal{H}}\left[ \sigma - Q_n \sigma Q_n\right]\nonumber\\
&=0,
\end{align}
and \eqref{eq:convergence-trace} follows.

To conclude that \eqref{eq:convergence-subplans} holds in the trace norm, define for $n$ sufficiently large the quantum state
\begin{equation}
\Pi_n = (Q_n \otimes P^T_n) \Pi (Q_n \otimes P^T_n) \left/ \mathrm{Tr}\left[ (Q_n \otimes P^T_n) \Pi (Q_n \otimes P^T_n)\right]\right.,
\end{equation}
so that $\lim_n \Pi_n = \Pi$ weakly, hence in the trace norm. By \eqref{eq:convergence-trace} this  implies convergence in the trace norm also in \eqref{eq:convergence-subplans}.
\end{proof}
\end{lem}

\subsection{Energy}\label{sec:energy}
We prove in this subsection some results that are needed to deal with the energy of quantum states with respect to unbounded operators.

\begin{lem}\label{lem:approx-unbounded}
Let $\mathcal{H}_A$, $\mathcal{H}_B$ be Hilbert spaces and let $A$ be a self-adjoint operator on $\mathcal{H}_A$. Then, there exists a sequence of bounded self-adjoint operators $\{A_n\}_{n=0}^\infty \subset \mathcal{B}(\mathcal{H})$ such that, for every for every quantum state $\rho \in \mathcal{S}(\mathcal{H}_A\otimes \mathcal{H}_B)$, $E_{A_n \otimes \mathbb{I}_B}(\rho)$ increases towards $E_{A \otimes \mathbb{I}_B}(\rho)$.
\begin{proof}
It is sufficient to consider the case of a pure state $\rho = |\psi\rangle\langle \psi |$, because the general case follows by \autoref{rem:en-pure} and the monotone convergence theorem for series.

Define by functional calculus \cite[Theorem VIII.5]{reed1980methods} $A_n = A \chi_{(-n, n)}(A)$, where $\chi_{(-n,n)}(x) = 1$ if $x \in (-n,n)$, $\chi_{(-n,n)}(x) = 0$ otherwise. Let us argue without the Hilbert space $\mathcal{H}_B$ first (\emph{i.e.}, $\mathcal{H}_B = \mathbb{C}$). Then,
\begin{equation}
E_{A_n}(\rho) = \left\|A_n|\psi\rangle\right\|^2 = \int_{-n}^n \lambda^2 \mathrm{d} \langle \psi | P_\lambda| \psi \rangle,
\end{equation}
where $P_\lambda  = \chi_{(-\infty, \lambda]}(A)$ denotes the resolution of the identity associated to $A$. The expression is increasing with respect to $n$ and $\sup_{n} \left\|A_n|\psi\rangle\right\|^2 < \infty$ if and only if $\psi$ belongs to the domain of $A$, because, defining $|\psi_n\rangle = \chi_{(-n,n)}(A) |\psi\rangle$, the sequence $\{ |\psi_n\rangle\}_{k=0}^\infty \subset \mathcal{H}$ converges towards $|\psi\rangle$ with $A |\psi_n\rangle$ bounded and $A$ is a closed operator. In this case, we also have $\lim_n A_n |\psi \rangle = A |\psi \rangle$ by \cite[Theorem VIII.5, c)]{reed1980methods}, hence the conclusion in this case. The thesis with the additional Hilbert space $\mathcal{H}_B$ follows from repeating the argument by noticing that
\begin{equation}
A_n \otimes \mathbb{I}_{B} = (A \otimes \mathbb{I}_B) \chi_{(-n,n)}(A \otimes \mathbb{I}_B).
\end{equation}
\end{proof}
\end{lem}

The following result shows that $E_A(\rho)$  does not depend on the ensemble that generates $\rho$.

\begin{prop}\label{lem:invariance-energy}
If $\rho = \sum_{k=0}^\infty |\phi_k\rangle\langle\phi_k|$ with $\{|\phi_k\rangle\}_{k=0}^\infty\subset \mathcal{H}$, then
\begin{equation}\label{eq:claimS}
E_A(\rho) = \sum_{k=0}^\infty \left\|A|\phi_k\rangle\right\|^2,
\end{equation}
where we let $\left\|A|\phi\rangle\right\|^2 =\infty$ if $| \phi\rangle$ does not belong to the domain of $A$.

Analogously, let $\mu$ be a probability measure on $\mathcal{H}$, and let
\begin{equation}
\rho = \int_\mathcal{H}|\psi\rangle\langle\psi|\,\mathrm{d}\mu(\psi)\,.
\end{equation}
Then,
\begin{equation}\label{eq:claimI}
E_A(\rho) = \int_\mathcal{H}\left\|A|\psi\rangle\right\|^2\mathrm{d}\mu(\psi)\,.
\end{equation}
\begin{proof}
If we define $|\hat \phi_k\rangle  = |\phi_k\rangle / \| |\phi_k\rangle\|$, then the thesis can be rewritten as
\begin{equation}
E_A(\rho) = \sum_{k=0}^\infty \| |\phi_k \rangle\|^2 E_A(|\hat \phi_k \rangle\langle \hat \phi_k | ),
\end{equation}
hence, by \autoref{lem:approx-unbounded} and the monotone convergence theorem for series, it is sufficient to prove the result for $A \in \mathcal{B}(\mathcal{H})$.

Arguing as in \cite[Theorem 2.6]{nielsen2010quantum} (noticing that the proof holds also in the infinite dimensional case), there exists a unitary  map   $U: \ell^2(\mathbb{N}) \to \ell^2(\mathbb{N})$,  $U^\dagger U = \mathbb{I}_{\ell(\mathbb{N})}$,  \emph{i.e.},
 $(u_{ij})_{i,j=0}^\infty$ with $\sum_{k=0}^\infty u_{ki}^* u_{kj} = \delta_{ij}$, such that
\begin{equation}
|\phi_k\rangle = \sum_{i=0}^\infty u_{ki} \sqrt{p_i} | \psi_i\rangle.
\end{equation}
Then,
\begin{align}
& \sum_{k=0}^\infty \left\|A|\phi_k\rangle\right\|^2  = \sum_{k=0}^\infty \langle \phi_k| A^2 | \phi_k\rangle  = \sum_{k=0}^\infty  \sum_{i,j=0}^\infty \sqrt{p_i p_j} u_{ki}^* u_{kj} \langle \psi_i| A^2 | \psi_j\rangle  \nonumber\\
&=  \sum_{i,j=0}^\infty \sqrt{p_i p_j}\langle \psi_i| A^2 | \psi_j\rangle \left( \sum_{k=0}^\infty u_{ki}^* u_{kj}\right) = \sum_{i=0}^\infty p_i \langle \psi_i| A^2 | \psi_i\rangle = E_A(\rho),
\end{align}
and the claim \eqref{eq:claimS} follows.

The integral version \eqref{eq:claimI} of the claim can be proved along the same lines employing Beppo Levi's monotone convergence theorem \cite[Theorem 2.8.2]{bogachev2007measure} and Fubini's theorem \cite[Theorem 3.4.4]{bogachev2007measure}.
\end{proof}
\end{prop}

\begin{prop}\label{prop:en}
$E_{A}$ is lower semicontinuous, \emph{i.e.}, if $\{\rho_n\}_{n=0}^\infty \subset \mathcal{S}(\mathcal{H})$ converge towards $\rho \in\mathcal{S}(\mathcal{H})$, then $E_A(\rho) \le \liminf_n E_A(\rho_n)$.
\begin{proof}
If $A$ is bounded then, by \autoref{rem:en}, $E_A(\rho) = \mathrm{Tr}_{\mathcal{H}} \left[A \, \rho \, A\right]$, which is continuous with respect to the trace norm. The general case follows by \autoref{lem:approx-unbounded}, since $E_A(\rho) = \sup_{n} E_{A_n}(\rho)$ and a pointwise supremum of continuous functions is lower semicontinuous.
\end{proof}
\end{prop}

\begin{prop}\label{prop:polarization}
Let $A$, $B$ be respectively self-adjoint operators on Hilbert spaces $\mathcal{H}_A$, $\mathcal{H}_B$ and let $\Pi \in \mathcal{S}(\mathcal{H}_A \otimes \mathcal{H}_B)$ with $\rho = \mathrm{Tr}_{A}[\Pi]$, $\sigma = \mathrm{Tr}_B[\Pi]$. Then
\begin{equation}\label{eq:energy-marginal}
E_{A \otimes \mathbb{I}_{B}} (\Pi) = E_{A}(\sigma), \quad  E_{\mathbb{I}_A \otimes B} (\Pi) = E_{B}(\sigma).
\end{equation}
If both energies above are finite, then
\begin{equation} \label{eq:energy-polarization}
E_{A \otimes \mathbb{I}_{B} + \mathbb{I}_A \otimes B} (\Pi)  + E_{A \otimes \mathbb{I}_{B} - \mathbb{I}_A \otimes B} (\Pi)= 2 E_{A}(\rho) + 2 E_{B}(\sigma).
\end{equation}
\begin{proof}
We prove the first identity in \eqref{eq:energy-marginal}, the second one being similar. By \autoref{lem:approx-unbounded}, it is sufficient to argue when $A$ is bounded. Then, by \autoref{rem:en},
\begin{align}
& E_{A \otimes \mathbb{I}_B}(\Pi) = \mathrm{Tr}_{AB} [ (A \otimes \mathbb{I}_B) \Pi (A \otimes \mathbb{I}_B)] \nonumber \\
& = \mathrm{Tr}_{A} [ A \mathrm{Tr}_B[ \Pi ] A ] = \mathrm{Tr}_{A}[A \sigma A] = E_{A}(\sigma).
\end{align}

To prove \eqref{eq:energy-polarization}, by \autoref{rem:en-pure}, is it sufficient to consider the case of a pure state $\Pi = |\psi\rangle \langle \psi |$, so that it reads
\begin{align}
&\|  \left( A \otimes \mathbb{I}_{B} + \mathbb{I}_A \otimes B \right) | \psi\rangle \|^2 + \| \left( A \otimes \mathbb{I}_{B} - \mathbb{I}_A \otimes B \right)| \psi\rangle\|^2\nonumber\\
&= 2 \|\left(A \otimes \mathbb{I}_{B} \right) | \psi\rangle\|^2 + 2 \| \left( \mathbb{I}_A \otimes B \right) | \psi\rangle\|^2
\end{align}
assuming that both terms in the right hand side are finite, \emph{i.e.},  $|\psi\rangle$ belongs to the domain of $A\otimes \mathbb{I}_B$ and $\mathbb{I}_A \otimes B$. But then
\begin{equation}
 \left( A \otimes \mathbb{I}_{B} + \mathbb{I}_A \otimes B \right) | \psi\rangle =  \left( A \otimes \mathbb{I}_{B} \right) | \psi\rangle+ \left( \mathbb{I}_A \otimes B \right)| \psi\rangle,
\end{equation}
\begin{equation}
 \left( A \otimes \mathbb{I}_{B} - \mathbb{I}_A \otimes B \right)| \psi\rangle =  \left(A \otimes \mathbb{I}_{B} \right) | \psi\rangle- \left( \mathbb{I}_A \otimes B \right) | \psi\rangle,
\end{equation}
hence the thesis by straightforward computation.
\end{proof}
\end{prop}

\subsection{Convergence of the quantum Wasserstein distance}\label{sec:CEproof}
As in \autoref{sec:W}, we fix quadratures $\{R_1,\,\ldots,\,R_N\}$, \emph{i.e.}, self-adjoint operators on $\mathcal{H}$. Then the cost associated to a quantum plan $\Pi\in \mathcal{C}(\rho, \sigma)$ (\autoref{def:cost}) reads
\begin{equation}\label{eq:c-sum-energies}
C(\Pi) = \sum_{i=1}^N E_{R_i\otimes\mathbb{I}_{\mathcal{H}^*} - \mathbb{I}_{\mathcal{H}}\otimes R_i^T}(\Pi)
\end{equation}
and a quantum state $\rho \in \mathcal{S}(\mathcal{H})$ has finite energy (\autoref{def:energy-main}) if
\begin{equation}
\sum_{i=1}^N E_{R_i}(\rho) <\infty\,.
\end{equation}

We are in a position to give a proof of \autoref{prop:CE}.

{\bf \autoref{prop:CE}}
\emph{Let $\rho,\,\sigma\in\mathcal{S}(\mathcal{H})$ have finite energy.
Then, any plan $\Pi\in\mathcal{C}(\rho,\sigma)$ has finite cost.}
\begin{proof}
For $i=1, \, \ldots, N$, we apply \autoref{prop:polarization} with $A = R_i$, $B = R_i^T$ so that
\begin{align}\label{eq:cost-polarization}
& C(\Pi)  = \sum_{i=1}^N E_{R_i\otimes\mathbb{I}_{\mathcal{H}^*} - \mathbb{I}_{\mathcal{H}}\otimes R_i^T}(\Pi) \nonumber \\
& = \sum_{i=1}^N\left( 2 E_{R_i}(\rho) + 2 E_{R_i}(\sigma) - E_{R_i\otimes\mathbb{I}_{\mathcal{H}^*} - \mathbb{I}_{\mathcal{H}}\otimes R_i^T}(\Pi)\right)< \infty,
\end{align}
where we also used that $E_{R_i^T}(\rho^T) = E_{R_i}(\rho)$.
\end{proof}

\begin{prop}\label{prop:continuity-plan}
Let $\rho$, $\sigma \in \mathcal{S}(\mathcal{H})$ be quantum states with finite energy and let $\{\Pi_n\}_{n=0}^\infty \subset \mathcal{S}(\mathcal{H}\otimes \mathcal{H}^*)$ be a sequence of quantum couplings $\Pi_n \in \mathcal{C}(\rho_n, \sigma_n)$ converging towards $\Pi \in \mathcal{C}(\rho, \sigma)$. Then $C(\Pi) \le \liminf_{n} C(\Pi_n)$. If moreover $\lim_{n}E_{R_i}(\rho_n) = E_{R_i}(\rho)$ and  $\lim_{n}E_{R_i}(\sigma_n)  = E_{R_i}(\sigma)$ for $i =1, \ldots, N$, then $C(\Pi) = \lim_n C(\Pi_n)$.
\begin{proof}
The first inequality  follows from \autoref{prop:en} applied to each term in \eqref{eq:c-sum-energies}.  Assuming convergence of the energies of both marginals, it is sufficient to show that that $C(\Pi) \ge \limsup_n C(\Pi_n)$. To this aim we use \eqref{eq:cost-polarization} and again \autoref{prop:en} to argue that
\begin{equation}
-E_{\left(R_i\otimes\mathbb{I}_{\mathcal{H}^*} + \mathbb{I}_{\mathcal{H}}\otimes R_i^T\right)}(\Pi_n) \ge \limsup_{n \to \infty} - E_{\left(R_i\otimes\mathbb{I}_{\mathcal{H}^*} + \mathbb{I}_{\mathcal{H}}\otimes R_i^T\right)}(\Pi_n).
\end{equation}
\end{proof}
\end{prop}

\begin{prop}\label{prop:existence-optimal-plan}
Let $\rho$, $\sigma \in \mathcal{S}(\mathcal{H})$ be quantum states with finite energy.
Then, there exists $\Pi \in \mathcal{C}(\rho, \sigma)$ such that $C(\Pi) = D(\rho, \sigma)^2$.
\begin{proof}
Let $\Pi_n \in \mathcal{C}(\rho, \sigma)$ be such that $\lim_n C(\Pi_n) = D(\rho, \sigma)^2$. By \autoref{lem:compactness-plans} we can assume, up to extracting a sub-sequence, that $\{\Pi_n\}_{n=0}^\infty$ converge towards $\Pi \in \mathcal{C}(\rho, \sigma)$. By \autoref{prop:continuity-plan}, $D(\rho, \sigma)^2 \le C(\Pi) \le \liminf_n C(\Pi_n) = D(\rho,\sigma)^2$, hence $\Pi$ is optimal.
\end{proof}
\end{prop}

\begin{thm}\label{thm:continuity-distance}
Let $\rho$, $\sigma \in \mathcal{S}(\mathcal{H})$ be quantum states with finite energy.
\begin{itemize}
\item If $\{\rho_n\}_{n=0}^\infty$, $\{ \sigma_n\}_{n=0}^\infty \subset \mathcal{S}(\mathcal{H})$ have finite energy and converge respectively towards $\rho$, $\sigma$, then
    \begin{equation}
    D(\rho, \sigma) \le \liminf_{n\to\infty} D(\rho_n, \sigma_n)\,.
    \end{equation}
\item Let $\{ P_n \}_{n=0}^\infty$, $\{ Q_n\}_{n=0}^\infty \subset \mathcal{B}(\mathcal{H})$ be such that $0 \le P_n \le \mathbb{I}_{\mathcal{H}}$, $0 \le Q_n \le \mathbb{I}_{\mathcal{H}}$, and
$\lim_n P_n\rho P_n = \rho$, $\lim_{n} Q_n\sigma Q_n = \sigma$ in the trace norm. Define, for $n$ sufficiently large,
\begin{equation} \rho_n = \frac{ P_n \rho P_n}{ \mathrm{Tr}[ P_n \rho P_n ]},\quad \sigma_n = \frac{ Q_n \sigma Q_n}{ \mathrm{Tr}[ Q_n \sigma Q_n]},
\end{equation}
and assume that  $\lim_{n}E_{R_i}(\rho_n) = E_{R_i}(\rho)$,  $\lim_{n}E_{R_i}(\sigma_n)  = E_{R_i}(\sigma)$  for  $i =1, \ldots, N$. Then
\begin{equation}
D(\rho, \sigma) = \lim_{n\to\infty} D(\rho_n, \sigma_n)\,.
\end{equation}
\end{itemize}
\begin{proof}
By \autoref{prop:existence-optimal-plan}, choose $\Pi_n \in \mathcal{C}(\rho_n, \sigma_n)$  such that
\begin{equation}
C(\Pi_n) = D(\rho_n, \sigma_n)^2\,.
\end{equation}
By \autoref{lem:compactness-plans} and \autoref{prop:continuity-plan} we obtain $\Pi \in \mathcal{C}(\rho, \sigma)$ such that  $C(\Pi) \le \liminf_n C(\Pi_n)$, hence
\begin{equation}
 D(\rho, \sigma)^2 \le C(\Pi) \le \liminf_n D(\rho_n, \sigma_n)^2.
 \end{equation}
 To prove the second statement, let $\Pi \in \mathcal{C}(\rho, \sigma)$ be such that $C(\Pi) = D(\rho, \sigma)^2$. Write
 \begin{align}
 \tilde{\Pi}_n & = (Q_n \otimes P_n^T) \Pi (Q_n \otimes P_n^T),\nonumber\\
  \tilde{\rho}_n^T & = \mathrm{Tr}_{\mathcal{H}}[\tilde{\Pi}_n] = P_n^T \mathrm{Tr}_{\mathcal{H}}[ (Q_n \otimes \mathbb{I}_{\mathcal{H}^*} ) \Pi  (Q_n \otimes \mathbb{I}_{\mathcal{H}^*}) ] P_n^T,\nonumber\\
    \tilde{\sigma}_n & =  \mathrm{Tr}_{\mathcal{H}^*}[\tilde{\Pi}_n] =  Q_n \mathrm{Tr}_{\mathcal{H}^*}[ (\mathbb{I}_{\mathcal{H}}\otimes P_n^T )\Pi  (\mathbb{I}_{\mathcal{H}} \otimes P_n^T) ] Q_n,\nonumber\\
    m_n &= \mathrm{Tr}_{\mathcal{H}\otimes \mathcal{H}^*} [\tilde{\Pi}_n] = \mathrm{Tr}_{\mathcal{H}^*}[\tilde{\rho}_n^T] = \mathrm{Tr}_{\mathcal{H}}[\tilde{\sigma}_n],
    \end{align}
and notice that $\tilde{\rho}_n^T \le P_n^T \rho^T P_n^T = \rho_n^T$ and $\tilde{\sigma}_n \le Q_n \sigma Q_n = \sigma_n$. We define
\begin{equation}
 \Pi_n = m_n \tilde{\Pi}_n + \tilde{\sigma}_n \otimes (\rho_n^T - \tilde{\rho}_n^T) +  (\sigma_n - \tilde{\sigma}_n) \otimes \tilde{\rho}_n^T + (\sigma_n - \tilde{\sigma}_n) \otimes (\rho_n^T - \tilde{\rho}_n^T).
 \end{equation}
It holds $\Pi_n \ge 0$,
\begin{equation} \mathrm{Tr}_\mathcal{H}[\Pi_n] = m_n \tilde{\rho}_n^T + m_n ( \rho_n^T - \tilde{\rho}_n^T) + (1-m_n) \tilde{\rho}_n^T + (1-m_n)(\rho_n^T - \tilde{\rho}_n^T) = \rho_n^T,
 \end{equation}
 and similarly $\mathrm{Tr}_\mathcal{H^*}[\Pi_n] =\sigma_n$, so that $\Pi_n \in \mathcal{C}(\rho_n, \sigma_n)$.

 We claim that $\lim_n \Pi_n = \Pi$. Since  $\lim_{n} P_n \rho P_n = \rho$ in the trace norm, $\lim_n\mathrm{Tr}[P_n \rho P_n] =  1$, so $\lim_n \rho_n =\rho$. Similarly, $\lim_n \sigma_n = \sigma$. \autoref{lem:convergence-subplans} gives $\lim_n \tilde{\Pi}_n = \Pi$ in the trace norm and, by continuity of partial trace, $\lim_n \tilde{\rho}_n  = \rho$, $\lim_n\tilde{\sigma}_n = \sigma$ and $\lim_n m_n = 1$, hence the claim is proved.

 By \autoref{prop:continuity-plan}, because the energies of the marginals converge, it follows that $C(\Pi) = \lim_n C(\Pi_n)$. We conclude that
 \begin{equation}
 D(\rho, \sigma)^2 = C(\Pi) \ge \limsup_{n\to\infty} D(\rho_n, \sigma_n)^2.
 \end{equation}
\end{proof}
\end{thm}

\begin{rem}\label{rem:finite-approx}
In \autoref{thm:continuity-distance} we may choose $P_n = \sum_{i=0}^n |\psi_i \rangle \langle \psi_i|$ where $\rho$ has the eigendecomposition $\rho = \sum_{k=1}^\infty p_i |\psi_i \rangle \langle \psi_i|$, and similarly for $Q_n$, by considering an eigendecomposition of $\sigma$. Indeed, by \autoref{lem:invariance-energy} it follows that $E_{R_i}(\rho_n)$ converge towards $E_{R_i}(\rho)$, and similarly for $\sigma$. We deduce that
\begin{equation}
\lim_{n\to\infty} D(\rho_n, \sigma_n) = D(\rho, \sigma)\,.
\end{equation}
\end{rem}

\section{}\label{sec:app}
\begin{lem}\label{lem:RX}
For any $X\in\mathcal{T}_2(\mathcal{H})$ and any self-adjoint operator $R$ on $\mathcal{H}$ such that $\left\|\left[R,\,X\right]\right\|_2<\infty$ we have
\begin{align}
\left\|\left[R,\,X\right]\right\|_2^2 &\ge \mathrm{Tr}_{\mathcal{H}}\left[R\left(X^\dag\,X + X\,X^\dag\right)R - \sqrt{X^\dag\,X}\,R\,\sqrt{X^\dag\,X}\,R\right.\nonumber\\
&\phantom{\ge \mathrm{Tr}_{\mathcal{H}}}\quad\left. - \sqrt{X\,X^\dag}\,R\,\sqrt{X\,X^\dag}\,R\right]\,.
\end{align}
\begin{proof}
Let us consider the singular-value decomposition of $X$:
\begin{equation}
X = \sum_{i=0}^\infty x_i\,|\psi_i\rangle\langle\phi_i|\,,\qquad x_i\ge0\,,\qquad \langle\psi_i|\psi_j\rangle = \langle\phi_i|\phi_j\rangle = \delta_{ij}\,,
\end{equation}
where the series converges in the Hilbert-Schmidt norm.
We get
\begin{align}
&\left\|\left[R,\,X\right]\right\|_2^2 = \mathrm{Tr}_{\mathcal{H}}\left[R\left(X^\dag\,X + X\,X^\dag\right)R - 2\,X^\dag\,R\,X\,R\right]\nonumber\\
&\ge \mathrm{Tr}_{\mathcal{H}}\left[R\left(X^\dag\,X + X\,X^\dag\right)R\right] - 2\left|\mathrm{Tr}_{\mathcal{H}}\left[X^\dag\,R\,X\,R\right]\right|\nonumber\\
&=\mathrm{Tr}_{\mathcal{H}}\left[R\left(X^\dag\,X + X\,X^\dag\right)R\right] -  2\left|\sum_{i,\,j=0}^\infty x_i\,x_j\,\langle\psi_i|R|\psi_j\rangle\langle\phi_j|R|\phi_i\rangle\right|\nonumber\\
&\ge \mathrm{Tr}_{\mathcal{H}}\left[R\left(X^\dag\,X + X\,X^\dag\right)R\right] -  \sum_{i,\,j=0}^\infty x_i\,x_j\left(\left|\langle\psi_i|R|\psi_j\rangle\right|^2 + \left|\langle\phi_j|R|\phi_i\rangle\right|^2\right)\nonumber\\
& = \mathrm{Tr}_{\mathcal{H}}\left[R\left(X^\dag\,X + X\,X^\dag\right)R - \sqrt{X^\dag\,X}\,R\,\sqrt{X^\dag\,X}\,R - \sqrt{X\,X^\dag}\,R\,\sqrt{X\,X^\dag}\,R\right]\,,
\end{align}
and the claim follows.
\end{proof}
\end{lem}

\bibliography{biblio}
\bibliographystyle{unsrt}

\end{document}